\documentclass[12pt]{article}
\usepackage[reqno]{amsmath}
\usepackage{epsfig}
\usepackage{array}
\usepackage{float}
\usepackage{lscape,graphicx}
\usepackage{amssymb}

\def\ltap{\ \raisebox{-.4ex}{\rlap{$\sim$}} \raisebox{.4ex}{$<$}\ }
\def\gtap{\ \raisebox{-.4ex}{\rlap{$\sim$}} \raisebox{.4ex}{$>$}\ }

\newcommand{\betabeta}{\mbox{$(\beta \beta)_{0 \nu}  $}}

\newcommand{\hbeta}{$\mbox{}^3 {\rm H}$ $\beta$-decay \ }

\newcommand{\bea}{\begin{equation}\begin{array}{c}}
\newcommand{\eea}{\end{array}\end{equation}}
\newcommand{\ea}{\end{array}} 

\newcommand{\beq}{\begin{equation}}
\newcommand{\eeq}{\end{equation}}
\newcommand{\bad}{\begin{array}{ccc}}

\newcommand{\ba}{\begin{array}{c}}
\newcommand{\N}{\mathbf{O}}
\newcommand{\I}{\mathbf{1}}
\newcommand{\half}{\frac{1}{2}}
\newcommand{\diag}{{\rm diag}}
\newcommand{\PMNS}{{\rm PMNS}}

\begin{document}
\hfill{{\small Ref. SISSA 36/2010/EP}}

\hfill{{\small Ref. TUM-HEP 763/10}}

\hfill{{\small Ref. IPPP/10/42, DCTP/10/84}}


\begin{center}
{\bf{\large TeV Scale See-Saw Mechanisms of Neutrino Mass Generation,}} 

{\bf{\large the Majorana Nature of the Heavy Singlet Neutrinos and 
$\betabeta$-Decay }}

\vspace{0.4cm}
A. Ibarra$\mbox{}^{a)}$,
E. Molinaro$\mbox{}^{b,c)}$
~and~
S. T. Petcov$\mbox{}^{b,d)}$
\footnote{Also at: Institute of Nuclear Research and
Nuclear Energy, Bulgarian Academy of Sciences, 1784 Sofia, Bulgaria}

\vspace{0.2cm}
$\mbox{}^{a)}${\em Physik-Department T30d, Technische Universit\"at M\"unchen,\\ James-Franck-Stra{\ss}e, 85748 Garching, Germany.\\}

\vspace{0.1cm}
$\mbox{}^{b)}${\em  SISSA and INFN-Sezione di Trieste, 
Via Bonomea 265, 34136 Trieste, Italy.\\}

\vspace{0.1cm}
$\mbox{}^{c)}${\em IPPP, Durham University, Durham DH1 3LE, UK.\\
}

\vspace{0.1cm}
$\mbox{}^{d)}${\em IPMU, University of Tokyo, Tokyo, Japan.\\
}
\end{center}

\begin{abstract}
It is shown that the Majorana nature of the heavy 
neutrinos $N_j$ having masses in the range 
of $M_j \sim (100 - 1000)$ GeV and present 
in the TeV scale type I
and inverse see-saw scenarios of 
neutrino mass generation, is unlikely to be observable 
in the currently operating and 
future planned accelerator experiments (including LHC)
due to the existence of very strong constraints 
on the parameters and couplings responsible for the 
corresponding $|\Delta L| = 2$ processes, 
$L$ being the total lepton charge.
If the heavy Majorana neutrinos $N_j$
are observed and they are associated only 
with the type I
or inverse see-saw mechanisms 
and no additional TeV scale ``new physics'', 
they will behave
like Dirac fermions to a relatively 
high level of precision, 
being actually pseudo-Dirac particles.
The observation of effects proving 
the Majorana nature of $N_j$ would imply that 
these heavy neutrinos
have additional relatively strong  
couplings to the 
Standard Model particles  
(as, e.g. in the type III see-saw scenario), 
or that light neutrino masses compatible with the 
observations are generated 
by a mechanism other than see-saw 
(e.g., radiatively at one or two loop level) 
in which the heavy Majorana 
neutrinos $N_j$ are nevertheless involved.

\end{abstract}

\section{Introduction}

\vspace{-0.6cm}
\hskip 1cm 

  The experiments with solar,
atmospheric, reactor and accelerator neutrinos
\cite{cleveland98,fukuda96,abd09,anselmann92,SKsolar02,ahmad01,
SKatm98,SKdip04,KL162,BOREX,ahn06,Michael06}
have provided compelling evidences for the
existence of flavour neutrino oscillations \cite{BPont57,MNS62} 
caused by nonzero neutrino masses and neutrino mixing.
These data imply the presence of neutrino mixing in the weak charged 
lepton current:
\begin{equation}
\nu_{l \mathrm{L}}(x)  = \sum_{j} U_{l j} \, \nu_{j \mathrm{L}}(x),~~
l  = e,\mu,\tau, 
\label{3numixCC}
\end{equation}
\noindent where 
$\nu_{lL}$ are the flavour neutrino fields, 
$\nu_{j \mathrm{L}}(x)$ is the 
left-handed (LH)
component of the field of 
the neutrino $\nu_j$ possessing a mass $m_j$ and $U$ 
is a unitary matrix - the
Pontecorvo-Maki-Nakagawa-Sakata (PMNS)
neutrino mixing matrix \cite{BPont57,MNS62,BPont67}. 

   All compelling  neutrino oscillation data
can be described assuming 
3-flavour neutrino mixing in vacuum.
The data on the invisible decay width of the $Z^0$-boson 
is compatible with only 3 light flavour neutrinos 
 coupled to $Z^0$ (see, e.g. \cite{Znu}). 
The number of massive neutrinos $\nu_j$, $n$, 
can, in general, be bigger than 3, $n>3$,
if, for instance, there exist right-handed (RH)
sterile neutrinos \cite{BPont67} 
and they mix with the LH flavour neutrinos. 
It follows from the existing data that 
at least 3 of the neutrinos $\nu_j$, say 
$\nu_1$, $\nu_2$, $\nu_3$, must be light,
$m_{1,2,3} \ltap 1$ eV, and must have different 
masses, $m_1\neq m_2 \neq m_3$. 
At present there are no compelling 
experimental evidences for the existence  
of more than 3 light neutrinos. 

  As is also well known, the data on the 
absolute scale of neutrino masses  
(including the data from  \hbeta experiments 
and astrophysical observations) 
imply that neutrino masses are much smaller than 
the masses of the charged leptons and quarks.
If we take as an indicative 
upper limit $m_j \ltap 0.5$ eV, we have 
$m_j/m_{l,q} \ltap 10^{-6}$, $l=e,\mu,\tau$, 
$q=d,s,b,u,c,t$. It is natural 
to suppose  that the remarkable smallness of neutrino 
masses is related to the existence of new 
fundamental mass scale in particle physics, 
and thus to new physics beyond that predicted 
by the Standard Model.

A natural explanation of the smallness of
neutrino masses is provided by the see-saw mechanism 
of neutrino mass generation \cite{seesaw}. An integral part of the
simplest version of this mechanism - the so-called 
``type I see-saw'', are the $SU(2)_L$ singlet 
RH neutrinos $\nu_{lR}$ (RH neutrino fields $\nu_{lR}(x)$). 
Within the see-saw framework, the latter are 
assumed to possess a Majorana mass term as well as 
Yukawa type coupling 
with the Standard Model 
lepton and Higgs doublets
$\psi_{lL}(x)$ and $\Phi(x)$, respectively, $l=e,\mu,\tau$.
In the basis in which the Majorana mass 
matrix of RH neutrinos is diagonal, 
the Majorana mass term of the RH neutrinos has 
the standard form $(1/2)\,M_{k}\overline{N_k}(x)\, N_k(x)$,
$N_k(x)$ being the heavy Majorana neutrino 
field possessing a mass $M_k > 0$. The fields
$N_k(x)$ satisfy the Majorana condition
$C \overline{N_k}^T(x) = \rho_k  N_k(x)$,
where $C$ is the charge conjugation matrix 
and $\rho_k$ is a phase.
When the electroweak 
symmetry is broken spontaneously, 
the neutrino Yukawa coupling generates a Dirac mass term:
$m^{D}_{li}\,\overline{\nu_{lL}}\,N_{iR}(x) + \hbox{h.c.}$,
with $m^{D} = v\lambda$, $\lambda_{li}$ being 
the matrix of neutrino Yukawa couplings 
and $v = 174$ GeV being the Higgs doublet v.e.v.
In the case when the elements of 
$m^{D}$ are much smaller than $M_k$,
$|m^{D}_{li}|\ll M_k$, $i,k=1,2,3$, $l=e,\mu,\tau$, 
the interplay between the
Dirac mass term and the Majorana mass term 
of the heavy Majorana neutrinos
$N_k$ generates an effective Majorana mass 
(term) for the LH flavour neutrinos (see, e.g. \cite{seesaw,BiPet87}):
$(m_{\nu})_{l'l}\cong -  m^{D}_{l'j}M^{-1}_{j}(m^{D})^T_{jl}$.
In grand unified theories, $m^{D}$ is typically of the
order of the charged fermion masses. In $SO(10)$ theories, 
for instance, $m^{D}$ coincides with the up-quark mass matrix.
Taking indicatively $m_{\nu} \sim 0.05$ eV, 
$m^{D}\sim 100$ GeV, one finds 
$M\sim 2\times 10^{14}$ GeV, which is close 
to the scale of unification of the electroweak and
strong interactions, $M_{GUT}\cong 2\times 10^{16}$ GeV.   
In GUT theories with RH neutrinos one finds 
that indeed the heavy Majorana neutrinos 
$N_j$ naturally obtain masses which are  
by few to several orders of magnitude smaller 
than $M_{GUT}$. 

  One can similarly obtain an effective Majorana mass term for the 
LH flavour neutrinos by introducing i) an $SU(2)_L$ triplet
of leptons, which includes a heavy neutral lepton and 
has an $SU(2)_L\times U(1)_{Y}$ invariant Yukawa coupling with 
the Standard Model Higgs doublet $\Phi(x)$
and the lepton doublets $\psi_{lL}(x)$
 (``type III see-saw mechanism'') \cite{Foot:1988aq}, 
or ii) by introducing additional neutral $SU(2)_L$ 
singlet fields $S_{\beta L}(x)$ which possess a 
Majorana mass term and couple to the RH singlet 
neutrino fields $\nu_{lR}$
(``inverse see-saw scenario'') \cite{Mohapatra:1986bd}.

  The estimate of $M_j$ given earlier is effectively 
based on the assumption that the neutrino 
Yukawa couplings are large: $|\lambda_{li}| \sim 1$.
The alternative possibility is to have heavy Majorana 
neutrino masses $M_j$ in the range of $\sim (100 - 1000)$ GeV, 
i.e. TeV scale see-saw generation of neutrino masses.
This possibility has received much attention recently 
(see, e.g. \cite{TeVsees}). One of the attractive features 
of the TeV scale see-saw scenarios is that the 
heavy Majorana neutrinos $N_j$ in such scenarios have 
relatively low masses which makes $N_j$ accessible 
in the experiments at LHC. This opens up the attractive 
prospect of having a see-saw 
mechanism of neutrino mass generation
which can be tested experimentally.

 One of the characteristic predictions of the 
type I, type III and the inverse see-saw models
is that both the light massive neutrinos and the heavy neutral 
neutrinos, which play a crucial role in these 
mechanisms, are Majorana particles.
The Majorana nature of the light neutrinos 
can be revealed in the neutrinoless double beta ($\betabeta$-) 
decay experiments (see, e.g. \cite{BiPet87,BPP1,bb0nudata}).
As was discussed by a large number of 
authors (see, e.g. \cite{HanPasc09} and the references quoted therein),
the Majorana nature of the heavy 
neutrinos of the TeV scale see-saw 
mechanisms can be established, in principle, 
in experiments at high energy accelerators, notably at LHC.  

  In the present article we revisit the low-energy neutrino physics 
constraints on the TeV scale type I 
and inverse see-saw models 
of neutrino mass generation. We concentrate on the constraints 
on the parameters of these models which are associated with the 
non-conservation of the total lepton charge $L$ and thus 
are directly related with the presence of light and heavy 
Majorana neutrinos in the indicated models.
We discuss the possibility to test the Majorana nature 
of the heavy 
Majorana neutrinos, which are an integral part 
on the indicated mechanisms of neutrino mass generation, 
at high energy accelerators, and in particular at LHC.

%
\section{See-Saw Scenarios with Two Mass Scales ($M_D$, $M_R$)}
%
\vspace{-0.8cm}
\hskip 1cm 

We consider, first, the standard type I see-saw 
scenario \cite{seesaw}, in which  we extend 
the Standard Model (SM) by adding 
$k$ ``heavy'' right-handed (RH) 
neutrino fields 
$\nu_{aR}$, $a=1,...,k$, $k\geq 2$.
We assume that the fields $\nu_{aR}$ are 
singlets with respect to the Standard Model 
gauge symmetry group, that they have 
Yukawa couplings with the left-handed (LH) 
lepton doublet fields and, in the spirit of the 
see-saw scenario, possess a ``large'' Majorana mass.
The neutrino mass term in the Lagrangian 
of the considered extension of the SM is given by:
\begin{equation}
\mathcal{L}_{\nu}\;=\; -\, \overline{\nu_{\ell L}}\,(M_{D})_{\ell a}\, \nu_{aR} - 
\half\, \overline{\nu^{C}_{aL}}\,(M_{N})_{ab}\,\nu_{bR}\;+\;{\rm h.c.}\,, 
\label{typeI}
\end{equation}
%
where $\nu^{C}_{aL}\equiv C \overline{\nu_{aR}}^T$, $C$ being the charge 
conjugation matrix, $M_{N} = (M_{N})^T$ is the 
$k\times k$ Majorana mass matrix of the RH neutrinos,
and $M_{D}$ is a $3\times k$ neutrino Dirac 
mass matrix which is generated by the matrix of 
neutrino Yukawa couplings after the electroweak 
(EW) symmetry breaking. The matrices $M_{N}$ and 
$M_{D}$ are complex, in general.
The full neutrino mass matrix 
in eq. (\ref{typeI}) can be set 
in a block diagonal form by the 
following transformation: 
\begin{eqnarray}
\Omega^T \left(\begin{array}{cc}
	\N & M_D \\
	M_D^T & M_N \\        
       \end{array}
 \right)\Omega & = &
\left(\begin{array}{cc}
	U^*\hat{m}U^\dagger & \N \\
	 \N^{T} & V^*\hat{M}V^\dagger \\        
       \end{array}
 \right) \label{seesaw}
\end{eqnarray}
%
where $\Omega$ is a $(3+k)\times (3+k)$ 
unitary matrix, $\hat{m}\equiv\diag(m_1,m_2,m_3)$ 
is a diagonal matrix with the masses of the light 
Majorana neutrinos, 
$\hat{M}\equiv\diag(M_1,M_2,\ldots,M_k)$ is a diagonal 
matrix containing the masses $M_{j}$ of the heavy 
Majorana neutrino mass eigenstates $N_{j}$ 
\footnote{The structure of the neutrino 
mass matrix shown in (\ref{seesaw}) 
appears also in type III see-saw scenario in which
the SM is extended by adding $k$ $SU(2)_{L}$ 
triplet fermion fields.}. 
The matrix $\N$ on the left-hand side of 
(\ref{seesaw}) is a $3\times 3$ matrix with all elements 
equal to zero. The same symbol is used on the right-hand 
side of (\ref{seesaw})  to indicate a $3\times k$ 
matrix with all null entries. The dimensions of 
the matrices $\N$ that appear in the 
 block mass matrix decompositions further in the text  
 will not be specified,
 but can similarly be easily deduced.
 
The unitary diagonalization matrix $\Omega$ can be 
formally expressed as the exponential of an 
antihermitian matrix:
\begin{eqnarray}
 \Omega & = & \exp\left(\begin{array}{cc}
  \N & R \\
  -R^\dagger & \N
\end{array}\right)\;=\;\left(
\begin{array}{cc}
\I-\half R R^\dagger & R \\
-R^\dagger & \I-\half R^\dagger R
\end{array}\right)\;+\;\mathcal{O}(R^3)\,,
\label{R0}
\end{eqnarray}
%
where  $R$ is a $3\times k$ complex matrix and
the second equality is obtained assuming that 
$R$ is ``small''. This assumption will be 
justified below.
In the case under discussion the PMNS 
\cite{BPont57,MNS62} neutrino mixing matrix  is given by :
\begin{equation}
 U_\PMNS \;=\; U_\ell^\dagger(\I+\eta) U \,,
 \label{PMNS}
\end{equation}
%
where
\begin{equation}
 \eta\;=\;-\half R R^\dagger\,,
 \label{eta}
\end{equation}
%
and  $U$ and $U_\ell$  diagonalise the 
  Majorana mass matrix $m_{\nu}$  of the LH flavour neutrinos and  
the charged lepton mass matrix $m_\ell$, respectively:
\begin{equation}
 U^{T}m_{\nu}U\;=\;\diag(m_{1},m_{2},m_{3})
\end{equation}
\begin{equation}
 U_\ell m_\ell m_\ell^\dagger U_\ell^\dagger \;=\; \diag(m_e^{2},m_\mu^{2},m_\tau^{2})
\end{equation}
%
 $m_{e}$, $m_{\mu}$ and $m_\tau$ being  the charged lepton 
masses.
The matrix $\eta$ parametrises the deviation from 
unitarity of the neutrino mixing matrix (\ref{PMNS}). 

 In what follows we will work in the basis in which the charged lepton 
mass matrix is diagonal 
\footnote{This can be done without loss of generality.}.
Accordingly, we set $U_\ell = {\bf 1 }$ in eq. (\ref{PMNS}).
The charged current (CC) and the neutral current (NC) 
weak interaction couplings involving the light Majorana 
neutrinos $\chi_j$ with definite mass $m_j$ have the form:
\begin{eqnarray}
\label{nuCC}
\mathcal{L}_{CC}^\nu 
&=& -\,\frac{g}{\sqrt{2}}\, 
\bar{\ell}\,\gamma_{\alpha}\,\nu_{\ell L}\,W^{\alpha}\;
+\; {\rm h.c.}
=\, -\,\frac{g}{\sqrt{2}}\, 
\bar{\ell}\,\gamma_{\alpha}\,
\left( (1+\eta)U \right)_{\ell i}\,\chi_{i L}\,W^{\alpha}\;
+\; {\rm h.c.}\,,\\
\label{nuNC} 
\mathcal{L}_{NC}^\nu &=& -\, \frac{g}{2 c_{w}}\,
\overline{\nu_{\ell L}}\,\gamma_{\alpha}\,
\nu_{\ell L}\,
Z^{\alpha}\;  
= -\,\frac{g}{2 c_{w}}\,
\overline{\chi_{i L}}\,\gamma_{\alpha}\,
\left (U^\dagger(1+\eta+\eta^\dagger)U\right)_{ij}\,\chi_{j L}\,
Z^{\alpha}\;.
\end{eqnarray}
 The charged current 
and the neutral current  
interactions of the heavy  Majorana fields 
$N_j$ with $W^\pm$ and $Z^0$ read:
\begin{eqnarray}
 \mathcal{L}_{CC}^N &=& -\,\frac{g}{2\sqrt{2}}\, 
\bar{\ell}\,\gamma_{\alpha}\,(RV)_{\ell k}(1 - \gamma_5)\,N_{k}\,W^{\alpha}\;
+\; {\rm h.c.}\,\label{NCC},\\
 \mathcal{L}_{NC}^N &=& -\frac{g}{2 c_{w}}\,
\overline{\nu_{\ell L}}\,\gamma_{\alpha}\,(RV)_{\ell k}\,N_{k L}\,Z^{\alpha}\;
+\; {\rm h.c.}\,\label{NNC}.
\end{eqnarray}
Therefore, independently of its origin, 
the mixing of the heavy (RH) Majorana neutrinos  
with the LH flavour neutrinos is
constrained by several low energy data, including 
$\betabeta$-decay  
\cite{del Aguila:2006dx,FernandezMartinez:2007ms,Abada:2007ux,Antusch:2008tz}. 
More specifically, the diagonal elements of $\eta$ 
are constrained taking into account
the lepton universality tests 
and the invisible decay width of 
the $Z^0$-boson, 
while upper bounds on the absolute values
of the off-diagonal elements of $\eta$ are obtained from
the existing experimental upper limits on the rates of the
radiative lepton decays, $\ell_{i}\to\ell_{j}+\gamma$. 
For singlet fields $N_j$ 
with masses above the EW symmetry breaking scale, i.e. bigger 
than $\sim$ 100 GeV, the resulting limits on the 
non-unitarity of the neutrino mixing matrix read 
\cite{Antusch:2008tz,Antusch:2006vwa}:
\begin{equation}
|\eta|\;<\;\left(
      \begin{array}{ccc}
       4.0\times 10^{-3} & 1.2\times 10^{-4} & 3.2\times 10^{-3}\\
       1.2\times 10^{-4} & 1.6\times 10^{-3} & 2.1\times 10^{-3}\\
       3.2\times 10^{-3} & 2.1\times 10^{-3} & 5.3\times 10^{-3}
      \end{array}\right)\,.\label{eta_bounds}
\end{equation}
%
The constraints given above 
allow to set upper bounds also on 
the couplings $RV$  of the heavy 
singlet fields $N_j$ with the Standard Model 
$W^{\pm}$ and charged leptons, and $Z^0$ and the 
LH active neutrinos (see (\ref{NCC}) and (\ref{NNC}), respectively).

 We will standardly assume further that $N_j$ 
have masses $M_j \gtap 100$ GeV and that $M_N$ is ``much bigger'' than
$M_D$. Using eq.  (\ref{R0})  and  the expression 
for the see-saw neutrino mass matrix  (\ref{seesaw}),   
we obtain the following relations at  leading order in $R$:
\begin{eqnarray}
 M_D\, -\,R^* M_N \; \simeq \; \N\,, && 
 \label{R2}\\
-\,M_{D}R^{\dagger} -\,R^{*}M_{D}^{T}  + R^{*}\,M_{N}\,R^{\dagger}  \simeq m_{\nu} =  U^*\hat{m}U^\dagger \,,&& 
\label{m} \\
M_N +R^T M_D+ M_D^T R - V^*\hat{M} V^\dagger \; \simeq \; \N \,.
\label{M} &&
\end{eqnarray}
%
Equation  (\ref{R2}) implies that  under assumptions made the matrix $R$ is indeed ``small'':
\begin{equation}
R^{*} \;\simeq\; M_D\,M_N^{-1}  \,. 
\label{R}
\end{equation}
%
We can express the light and heavy neutrino mass
matrices in (\ref{m}) and (\ref{M}) in terms of $M_N$ and $R$:
\begin{eqnarray}
m_{\nu} &\equiv& U^*\hat{m}U^\dagger \; = \; -R^* M_N R^\dagger \,, 
\label{m2}  \\
V^*\hat{M} V^\dagger & \simeq & M_N +R^T R^* M_N+M_N R^\dagger R\,. 
\label{M2} 
\end{eqnarray}
%
The usual type I see-saw expression for the Majorana 
mass matrix of the  LH flavour neutrinos  
is easily recovered from eqs. (\ref{R})
and (\ref{m2}):  $m_{\nu}\simeq -M_{D} M_N ^{-1}M_{D}^{T}$. 

  In the basis we choose to work and up to 
corrections $\propto  R R^\dagger$,
the elements of the light neutrino mass 
matrix $m_{\nu}$ are given by:
$m_{\nu} \equiv U^*\hat{m}U^\dagger  \cong  U_{PMNS}^*\hat{m}U_{PMNS}^\dagger$. 
Using the existing upper limits on the absolute scale of neutrino masses
and the data on the neutrino mixing angles, obtained in neutrino oscillation 
experiments, one can derive the ranges of possible values of the elements 
of $m_{\nu}$  \cite{Merle:2006du}. 
For the purpose of the 
present study it is sufficient to use the approximate 
upper bounds 
$|(m_{\nu})_{l'l}| \lesssim 1$ eV, $l,l'=e,\mu,\tau$. 
From eqs. (\ref{m2}) and (\ref{M2}) 
we obtain to leading order in $R$:
\begin{equation}
\sum_{k} |(RV)^*_{l'k}\;M_k\, (RV)^{\dagger}_{kl}| \lesssim 1~{\rm eV}\,,
~l',l=e,\mu,\tau\,.
\label{VR1}
\end{equation}
%

  In the case of the element $(m_{\nu})_{ee}$, the bound 
follows from the experimental data on the  neutrinoless double beta 
($\betabeta$-) decay \cite{bb0nudata}. In this case, in addition 
to the standard contribution due to the light Majorana 
neutrino exchange, the  $\betabeta$-decay effective 
Majorana mass $(m_{\nu})_{ee}$ (see, e.g. \cite{BiPet87,BPP1}) 
receives a contribution from the exchange of the heavy Majorana 
neutrinos $N_k$. Taking into account this 
contribution as well, we get \cite{HPR83,JV83,HaxStev84}:
\begin{equation}
|(m_{\nu})_{ee}| \cong 
\left |\sum_{i}(U_\PMNS)^2_{ei}\, m_i 
- \sum_k\, F(A,M_k)\, (RV)^2_{e k}\,M_k \right |\,,
\label{mee1}
\end{equation}
%
where $F(A,M_k)$ is a known real (positive) function
of the atomic number $A$ of the decaying nucleus
and of the mass $M_k$ of $N_k$ \cite{HPR83,JV83,HaxStev84}.
Using the fact that 
$U_{PMNS}\,\hat{m}\,U_{PMNS}^T \cong  U\,\hat{m}\,U^T$
and eqs.  (\ref{m2}) and ( \ref{M2}), we obtain:
\begin{equation}
|(m_{\nu})_{ee}| \cong 
\left |\sum_{k}(RV)^2_{e k}\,M_k \left (1  + F(A,M_k)\right) \right|\,.
\label{mee1}
\end{equation}
%
The function $F(A,M_k)$ exhibits a rather 
weak dependence on $A$, which for the purpose of the 
present discussion can be neglected, and a relatively strong 
dependence on $M_k$. For $M_k = 100~(1000)$ GeV, 
an estimate of the largest possible values of $F(A,M_k)$
gives (see, e.g. \cite{Blen10}):
$F(A,M_k) \cong 7\times 10^{-6}~(7\times 10^{-8})$.
Clearly, in the case of interest the contribution due 
to the exchange of the heavy Majorana neutrinos in 
$(m_{\nu})_{ee}$ is subdominant and can be neglected.
This contribution can be relevant if, for instance, 
the ``leading order'' term is very strongly suppressed 
or if $\sum_{k}(RV)^2_{e k}\,M_k  = 0$.

   Using the upper bounds in eq. (\ref{VR1}) 
and barring ``accidental'' cancellations or 
extreme fine-tuning (at the level of $\sim 10^{9}$, see, 
e.g. \cite{Kersten:2007vk,Xing:2009ce}),
we get for the heavy Majorana neutrinos  $N_k$  
having masses $M_k \sim M_R \geq 100$ GeV 
the well-known strong constraint 
on the couplings of 
$N_k$ to the weak $W^{\pm}$ and $Z^0$ bosons
and charged leptons and light neutrinos:
\begin{equation}
 | (RV)_{l k}|\; \lesssim 3\times10^{-6}\left(\frac{100\; {\rm GeV}}{M_R}\right)^{1/2}\,,~~l=e,\mu,\tau,~j=1,2,...,k\,.
\end{equation}
This constraint 
\footnote{In principle, one can obtain a more 
refined constraint on $|(RV)_{e k}|$ using 
the existing limits on $|(m_{\nu})_{ee}|$ 
(see, e.g. \cite{bb0nudata}). 
However, the approximate upper bound of 1 eV we are using 
is sufficient for the purposes of the present discussion.}
makes the heavy Majorana neutrinos $N_j$ practically 
unobservable even at LHC (see, e.g. \cite{HanPasc09}).

   In order for the CC and NC couplings of the heavy Majorana 
neutrinos $N_j$ to $W^\pm$ and $Z^0$, 
eqs. (\ref{NCC}) and (\ref{NNC}), to be sufficiently large
so that the see-saw mechanism could 
be partially or completely tested in experiments at 
the currently operating and planned future accelerators 
(LHC included), the suppression 
implied by the inequality (\ref{VR1}) should be 
due to strong mutual compensation between the terms 
in the sum in the left-hand side of (\ref{VR1}).
Such cancellations arise naturally from 
symmetries in the lepton sector, 
corresponding, e.g. to the conservation 
of some additive lepton charge $\hat{L}$ 
(see, e.g. \cite{LeungSTP83,LWDW83,BiPet87,Branco:1988ex,Amit10}).
However, in the exact symmetry limit 
in this case the heavy neutrinos with 
definite mass and relatively large couplings 
to the $W^{\pm}$ and $Z^0$
should be Dirac particles, which is possible 
for all heavy neutrinos only if the number of the 
RH singlet neutrino fields $k$ is even: 
$k= 2q$, $q=1,2,..$. If their number is odd, 
barring again ``accidental'' cancellations
some (odd number) of the discussed 
heavy Majorana neutrinos will have
strongly suppressed couplings 
to the $W^{\pm}$ and charged leptons
and will be practically unobservable in 
the current and the future planned 
accelerator experiments.
Further, 
the spectrum of masses 
of the three light neutrinos, 
which depends on the assumed symmetry, 
typically would not correspond to the 
observations. The correct light Majorana 
neutrino mass spectrum 
can be generated by small perturbations 
that violate the corresponding symmetry, 
leading to the non-conservation of the 
lepton charge $\hat{L}$.
These perturbations split each heavy 
Dirac neutrino into two heavy Majorana 
neutrinos with close but different masses,
i.e. the heavy Dirac neutrinos become 
heavy pseudo-Dirac neutrinos \cite{LW81,STPPD82}.
The perturbations will have practically 
negligible effect on the couplings 
of the heavy Dirac states to the 
 $W^\pm$ and $Z^0$.
If, for instance, 
$|(RV)_{l j}|\sim 10^{-3}~(10^{-4})$,
the splitting between the masses of the 
two heavy Majorana neutrinos 
forming a pseudo-Dirac pair, 
as it follows from eq. (\ref{VR1}),
should satisfy roughly
$|\Delta M_{PD}| \lesssim 1~(100)$ MeV 
for masses of the order of 100 (1000) GeV.
Thus, the effect of the perturbations 
on the low-energy phenomenology of 
the indicated heavy neutrino 
states will be essentially negligible and 
to a high level of precision they 
will behave like Dirac fermions
\footnote{For the signatures of production 
of such TeV scale pseudo-Dirac neutrinos 
at LHC see, e.g. \cite{HanPasc09,delAguila:2008hw}.}.

  The preceding discussion implies that
the Majorana nature of the heavy Majorana 
neutrinos of the type I see-saw mechanism will be 
unobservable at LHC and the planned future
accelerator experiments. If heavy neutrinos are 
observed and they are associated with the 
type I see-saw mechanism without any additional 
TeV scale ``new physics'' (e.g. in the form of $Z'$ boson 
associated with an additional $U(1)$ local 
gauge symmetry, see, e.g. \cite{SKhalil10}
and references quoted therein),
they will behave like Dirac fermions
to a relatively high level of precision. 
The observation of effects proving 
their Majorana nature would imply that 
these heavy neutral leptons 
have additional relatively strong 
non-Standard Model couplings to the 
Standard Model particles, 
or that $M_DM^{-1}_NM^T_D \cong 0$ and 
$m_{\nu} \neq 0$ compatible with the observations 
arises as one and/or two loop higher order correction 
(see, e.g. \cite{SPTosh84,Amit10}).

  We will illustrate some of these 
conclusions/considerations 
with few simple examples.\\

{\bf The Case of a Broken Symmetry.} 
We will consider first the case when the Majorana 
mass matrix for the LH flavour neutrinos 
$m_{\nu} \neq 0$ arises as a result of breaking of a 
global symmetry corresponding to 
the conservation of a lepton charge.
In the symmetry limit one has $m_{\nu} = 0$.

Suppose we have two LH flavour neutrino 
fields $\nu_{lL}$, $l=e,\mu$, 
and two RH neutrino fields $\nu_{aR}$, $a=1,2$.
Let us assign a lepton charge $L_a$ to each
of the two RH neutrino fields:
$L_{a}(\nu_{bR}) = -\delta_{ab}$, 
i.e. $\nu_{1R}$ has lepton charge 
$L_1 =-1$ and lepton charges $L_{2} = L_l = 0$,  
$l=e,\mu$. Suppose that the Majorana mass 
matrix $M_N$ in eq. (\ref{typeI}) has the form:
\begin{eqnarray}
M_N = \left(
     \begin{array}{cc} 
   0 & M_{12} \\
   M_{12} & 0 
\end{array}\right)\,.
\label{MN2}
\end{eqnarray}
%
We take (for concreteness) 
$M_{12}$ to be real and positive,
$M_{12} > 0$, and assume also that 
$M_{12}\geq 100$ GeV. The motivation 
for the choice made of the form of $M_N$ 
is that, as is well known,
the matrix in eq. (\ref{MN2}) has two 
eigenvalues which have equal absolute values 
but opposite signs. Thus, one can expect 
that this may lead to the requisite suppression 
of the sum in the left-hand side of eq. (\ref{VR1}). 

  The Majorana mass 
term of the RH neutrinos 
in eq. (\ref{typeI}) with the mass matrix 
$M_N$ given by eq. (\ref{MN2}) 
conserves the lepton charge 
$(L_1 - L_2)$. It is diagonalised 
with the help of a $2\times 2$ orthogonal 
matrix ${\bf V}(\theta)$ with $\theta = \pi/4$.
The heavy Majorana mass-eigenstates $N_{1,2}$
have masses $M_{1} = M_{2} = M_{12} \equiv M$ and 
satisfy the Majorana conditions 
$C \overline{N_{k}}^T =\rho_k N_k$, 
$k=1,2$, where $\rho_1 = -1$ and 
$\rho_2 = +1$ ~
\footnote{The difference in the sign factors 
in the Majorana conditions for $N_{1,2}$ which 
have positive masses,
reflects the difference in the signs of the 
two eigenvalues of the matrix $M_N$, 
eq. (\ref{MN2}) (for a more detailed discussion 
see \cite{BiPet87}).}.
They are equivalent 
to one heavy Dirac neutrino field 
$N_D = (N_{1} + N_{2})/\sqrt{2}$ 
having a mass $M$.
The relations between the fields 
$\nu_{aR}$, $\nu^{C}_{aL}$,
$a=1,2$, and the fields $N_{k L}$, $N_{k R}$, 
$k=1,2$, and $N_{D L}$, $N_{D R}$,
$C \overline{N_{D L}}^T \equiv N^C_{D R}$ and
$C \overline{N_{D R}}^T \equiv N^C_{D L}$ 
have the form:
\begin{eqnarray}
\nu_{1R}&=&N_{D R} = \frac{1}{\sqrt{2}}\left (N_{1R} + N_{2R} \right)\,,~~
\nu_{2R} 
= N^C_{D R} = \frac{1}{\sqrt{2}}\left (-\,N_{1R} + N_{2R} \right)\,,
\label{nuR}\\[0.25cm]
\nu^{C}_{1L}&=&N^C_{D L} = \frac{1}{\sqrt{2}}\left (-\,N_{1L} + N_{2L} \right)\,, ~~  
\nu^{C}_{2L} 
=  N_{D L} = \frac{1}{\sqrt{2}}\left (N_{1L} + N_{2L} \right)\,.
\label{nuCL} 
\end{eqnarray}
%
Let us denote the elements of the Dirac mass matrix 
$M_D$ (see eq. (\ref{typeI})) as $(M_D)_{lk} \equiv m^D_{lk}$,
$l=e,\mu$, $k=1,2$. We will assume for simplicity that 
$M_D$ is a real matrix and that 
$|m^D_{lk}| \ll M_{12}$. The Majorana mass matrix for 
the LH flavour neutrinos, generated by the 
see-saw mechanism, has the form:
\begin{eqnarray}
 m_{\nu} & = & -R^* M_N R^\dagger \simeq -M_{D} M_N ^{-1}M_{D}^{T} \\
& = & -\frac{1}{M} \left(\begin{array}{cc}
  2m^D_{e1}m^D_{e2}  & m^D_{e1}m^D_{\mu 2} + m^D_{e2} m^D_{\mu 1}\\
   m^D_{e1}m^D_{\mu 2} + m^D_{e2} m^D_{\mu 1} &  2m^D_{\mu 1}m^D_{\mu 2}
\end{array}\right)\;.
\label{mnu2nu}
\end{eqnarray}
%
The matrix of CC and NC couplings of the heavy 
Majorana neutrinos $N_{1,2}$ to the 
SM $W^{\pm}$ and $Z^0$ bosons  reads:
\begin{eqnarray}
RV & = &\frac{1}{\sqrt{2}}\frac{1}{M}\left(\begin{array}{cc}
  m^D_{e2} - m^D_{e1}  & m^D_{e1} + m^D_{e2}\\
   m^D_{\mu 2} - m^D_{\mu 1} &  m^D_{\mu 1} + m^D_{\mu 2}
\end{array}\right)\;.
\label{RV2nu}
\end{eqnarray}
%
The form of $m_{\nu}$, eq. (\ref{mnu2nu}), reflects the 
the fact that  the contributions due to $N_1$ and $N_2$ 
tend to cancel each other: we have,  for instance,  
$(m_{\nu})_{ee} = -[(m^D_{e2} + m^D_{e1})^2  - (m^D_{e1} - m^D_{e2})^2]/(2M)$, 
etc.

  With  $|(m_{\nu})_{l'l}| \ltap 1$ eV and $M\geq 100$ GeV
we indeed get, in general, $|(RV)_{lk}| \ltap 10^{-6}$, 
$l=e,\mu$, $k=1,2$. The constraints under 
discussion on some of the elements of 
the matrix $(RV)$ (and thus on some of the 
couplings of $N_{1,2}$ to the 
$W^{\pm}$ and $Z^0$ bosons) can be 
avoided if some of the elements of 
the Dirac mass matrix are sufficiently small, 
so that $|(m_{\nu})_{l'l}| \ltap 1$ eV 
is satisfied, and at the same time 
not all elements of  $(RV)$ are 
suppressed. This possibility can be realised 
if, for instance,  $m^D_{e 1}/M$ and  $m^D_{\mu 1}/M$
are sufficiently small. We will set them to 
zero in what follows. In this limit 
we have $|(m_{\nu})_{l'l}| = 0$. The couplings  
$m^D_{e 2}/M$ and  $m^D_{\mu 2}/M$ are not 
constraint, except by the assumption
that $|m^D_{l 2}|/M \ll 1$, $l=e,\mu$.

 If $m^D_{e 1}/M = m^D_{\mu 1}/M = 0$,
the heavy Majorana neutrinos $N_{1,2}$ 
couple to the weak $W^{\pm}$ and $Z^0$ 
bosons only in the combination 
$(N_{1L} + N_{2L})/\sqrt{2} = N_{DL}$, 
i.e. only through the LH component of the 
Dirac field $N_D$. Moreover, in this case 
there is a conserved lepton charge 
$\hat{L} = L_e + L_{\mu} + (L_1 - L_2)$.  
This implies that \cite{BiPet87,LeungSTP83} 
the theory contains one heavy Dirac neutrino $N_D$
and two massless neutrinos $n_{1,2}$. The massless 
neutrino fields $n_{1L}$ and $n_{2L}$
are the dominant components of the LH 
flavour neutrino fields $\nu_{lL}$, $l=e,\mu$,
while the $N_{DR}$ and   $N^C_{DR}$
are the dominant components of the 
two RH neutrino fields, $\nu_{aR}$, $a=1,2$.
For the Majorana mass matrix of the LH flavour 
neutrinos we have in this approximation:    
$m_{\nu} = 0$.

  It should be clear that 
in the approximation being discussed,
there are no physical (observable) effects 
associated with the fact that 
$N_{1,2}$ are Majorana particles: 
$N_{1,2}$ always appear in the interaction 
Lagrangian in the combination 
$(N_{1L} + N_{2L})/\sqrt{2}$ which 
is equivalent to a Dirac fermion.  
The probability of having, e.g. same sign 
dilepton events 
\footnote{One of the charged 
leptons (say $\mu^-$) is produced together 
with $N_{1,2}$, while the second ($\mu^-$) 
is supposed to originate from the
$N_{1,2}$ decay into $W^\pm$ + charged lepton
($W^+ + \mu^-$).}, which would be a signature 
of the Majorana nature of $N_{1,2}$, is zero.
This is a consequence of the fact that 
the contributions of $N_{1}$ and $N_{2}$ 
in the amplitudes of the processes of
same sign dilepton production 
are equal in absolute value, but have opposite
signs and cancel completely each other.

 Indeed, consider the process of same sign di-muon 
production in $p-p$ collisions, assuming that 
one of the muons, say $\mu^-$, is produced together 
with real or virtual $N_{1,2}$, while the second $\mu^-$ 
originates from  the decay $N_{1,2} \rightarrow W^+ + \mu^-$,
with virtual or real $W^+$ (see, e.g. \cite{HanPasc09}). 
The  $W^+$ decays further
into, e.g. two hadron jets. If
in the general case of $M_1 \neq M_2$,
$M_1 < M_2$, the heavy Majorana 
neutrino $N_1$ is real, 
the invariant mass of the two 
jets and the second muon should be equal to 
the mass of $N_1$. Note that $N_{1,2}$ are not 
directly detected. The observation of this process
with the characteristic Breit-Wigner enhancement
of the cross section due to the $N_1$ propagator
when the invariant mass of 
the two jets and the second muon approaches 
the mass of $N_1$, would be a signature 
of the Majorana nature of $N_{1}$.
The part of the amplitude of the process 
under discussion, which is of interest for the purposes 
of the present analysis, has the form: 
\begin{equation}
P_{1,2} = \frac{(m^D_{\mu 2})^2}{M^2_{12}}\left [
 \frac{s^2 M_2}{p^2 - M^2_2 + i\Gamma_2 M_2}\, 
-\, \frac{c^2 M_1}{p^2 - M^2_1 + i\Gamma_1 M_1} \right ]\,,
\label{ppmumu12}
\end{equation}
%
where $p$ is the four momentum of the 
real (or virtual) $N_{1,2}$, $\Gamma_{1,2}$ is 
the width of $N_{1,2}$,
$c^2 = \cos^2\theta$,  $s^2 = \sin^2\theta$, 
where $\theta$ is the angle in 
$V$ which diagonalises the RH neutrino mass matrix. 
In deriving eq. (\ref{ppmumu12}) we have taken 
into account the fact that the sign factors 
in the Majorana conditions for 
$N_{1}$ and $N_{2}$ are opposite:
$\rho_1 = -1$ and $\rho_2 = +1$.
For the width of $N_{1,2}$ one has
for the ranges of masses of $N_{1,2}$ of interest
\cite{HanPasc09}:
$\Gamma_{1(2)} \propto G_F\, M^3_{1(2)} \ll M_{1(2)}$.
We note that $p^2$ is equal to the square of the 
invariant mass of the two jets and the second muon
in the final state of the process 
(see, e.g. \cite{HanPasc09}).
The above expression is valid for any value of 
the invariant mass $p^2$. It should be emphasised that 
even when $N_{1}$ is on mass shell, $p^2 = M^2_{1}$,
and the second term in eq. (\ref{ppmumu12}) dominates 
due to the  fact the $M_2$ is significantly 
bigger than $M_1$, the contribution of the virtual 
$N_2$ (i.e., the first term in eq. (\ref{ppmumu12})) 
is always present in the amplitude. 
In the specific case we are considering 
one has actually
$M_1 = M_2$, $\Gamma_{1}= \Gamma_{2}$,  
$c^2 = s^2 = 1/2$, and therefore 
$P_{1,2} =0$, as was suggested earlier.

 Consider next the ``perturbation'' of 
the scheme discussed by having, e.g. 
$m^D_{\mu 1} = 0$ but  $m^D_{e 1} \neq 0$. 
In this case $\hat{L}$ is no 
longer conserved: there is no 
conserved lepton charge in the theory. 
Correspondingly, $m_{\nu} \neq 0$ 
(except for the element  $(m_{\nu})_{\mu\mu} = 0$).
The two heavy Majorana neutrinos $N_{1,2}$
have slightly different masses now, 
$|M_{1}- M_{2}| \cong 
2|m^D_{e 1}m^D_{e 2}|/M $, 
forming a pseudo-Dirac neutrino 
$N_{PD} = (N_{1} + N_{2})/\sqrt{2}$.
\cite{LW81,STPPD82}.
The light neutrinos $n_{1,2}$ have nonzero masses 
$m_{1,2} \cong m^D_{e 1} [ \sqrt{(m^D_{e 2})^2 
+ (m^D_{\mu 2})^2 } \mp m^D_{e 2}]/M$. 

 If we assume that $|m^D_{e 1} |$ is of the same order 
as $|m^D_{e 2}|$ and  $|m^D_{\mu 2}|$,
the constraint (\ref{VR1}) applies and $N_{1,2}$ 
would be hardly observable, e.g. at LHC.
If, however, $m^D_{e 1} \neq 0$ is generated 
as a small perturbation, i.e. if 
$|m^D_{e 1}| \ll |m^D_{e 2}|, |m^D_{\mu 2}|$, one can have 
$(m_{\nu})_{l'l} \lesssim 1$ eV for relatively large 
$|m^D_{e 2}|/M$ and/or  $|m^D_{\mu 2}|/M$ 
couplings of $N_{1,2}$ to the $W^{\pm}$ and $Z^0$ bosons. 
This would make possible the production 
of $N_{1,2}$ with observable rates at LHC. 
However, also in this case 
$N_1$ and $N_2$ couple to $W^{\pm}$ and $Z^0$  
only in the combination $(N_{1L} + N_{2L})/\sqrt{2}$.
Moreover, they form a pseudo-Dirac neutrino 
with an extremely small mass splitting. 
Indeed, if for instance, 
$|m^D_{e 2}|,|m^D_{\mu 2}|\cong 10^{-3}~M$ 
and $M = 100$ GeV, using $|m_{1,2}| \ltap 1$ eV we get 
$|m^D_{e 1}|\ltap 1$ keV and 
$|M_{1}- M_{2}| \ltap 1$ eV. Actually, we have 
$|M_{1}- M_{2}| \cong |m_2 - m_1|$.
Thus,  $N_{1}$ and  $N_{2}$ form a pseudo-Dirac neutrino
which, given the tiny mass splitting between $N_{1}$ and  $N_{2}$,
will behave for all practical purposes as a Dirac neutrino.
The magnitude of all effects 
revealing the Majorana nature of the heavy neutral leptons $N_{1,2}$
is proportional to their mass difference, i.e. to the  factor 
$|m^D_{e 1} m^D_{e 2}|/M^2$, which renders these effects 
unobservable (e.g. at LHC). 

 We will extend next the previous rather straightforward
analysis to the case of 3 LH flavour neutrinos 
and three RH neutrinos $\nu_{aR}$, $a=1,2,3$. 
Consider for simplicity a model in which 
one of the three light Majorana neutrinos is massless:
\begin{equation}
	M_D = \left(
     \begin{array}{ccc} 
  0 & m^{D}_{e2} & m^{D}_{e3} \\
     0   & m^{D}_{\mu 2} & m^{D}_{\mu3}\\
     0   & m^{D}_{\tau 2} & m^{D}_{\tau 3}
\end{array}\right)\,,\;\;\;\;\;\;\;\;
M_N = \left(
     \begin{array}{ccc} 
   M_{11} & 0 & 0 \\
     0   & 0 & M_{23}\\
     0   & M_{23} & 0
\end{array}\right)\,.
\end{equation}
%
As a consequence of the simplifying 
choice made $m^{D}_{l1} = 0$, $l=e,\mu,\tau$,
the field $N_{1} = (\nu_{1R} + \nu^{C}_{1L})/\sqrt{2}$ is decoupled.
In the limit $m^{D}_{\ell 2 }=0$, $\ell= e,\mu,\tau$, there is a conserved 
lepton charge:  
$L^{\prime}\equiv L_{e}+L_{\mu}+L_{\tau}+L_{3}-L_{2}$. 
In this case the theory contains three massless
and one massive Dirac neutrinos \cite{LeungSTP83} 
(see also \cite{BiPet87}). 
The three massless neutrinos are the dominant 
components of the three LH flavour neutrinos.
The massive Dirac neutrino $N_D = (N_{2} + N_{3})/\sqrt{2}$
has a mass $M = M_{23} > 0$, 
where the two heavy Majorana neutrinos 
$N_{2}$ and $N_{3}$ have the same mass 
$M_{2} = M_{3} = M_{23}$  
and satisfy the Majorana conditions   
$C \overline{N_{k}}^T =\rho_k N_k$, 
$k=2,3$, where $\rho_2 = -1$ and $\rho_3 = +1$.
It should be clear that in this case there are no 
observable effects associated with the Majorana nature of the
$N_{2}$ and $N_3$.

 Consider next the case of  $m^{D}_{\ell 2,3 } \neq 0$, $\ell= e,\mu,\tau$.
Now there is no conserved lepton charge and
the resulting light neutrino mass spectrum has the form:
\begin{equation}
 m_{1} = 0\,,\;\;\;\; 
m_{2}\cong\frac{1}{M_{23}}\left(\sqrt{\Delta}  - A\right)\,,
\;\;\;\;m_{3}\cong\frac{1}{M_{23}}\left(\sqrt{\Delta}  + A\right)\,,
 \end{equation}
%
where
\begin{equation}
\Delta = \left( m^{D\, 2}_{e2} + m^{D\, 2}_{\mu 2}+m^{D\, 2}_{\tau 2}\right)\left( m^{D\, 2}_{e3} + m^{D\, 2}_{\mu 3}+m^{D\, 2}_{\tau 3}\right)\,,
\end{equation}
%
and 
\begin{equation}
 A =  m^{D}_{e2}m^{D}_{e3}+m^{D}_{\mu 2} m^{D}_{\mu 3}
+ m^{D}_{\tau 2} m^{D}_{\tau 3}\,.
\end{equation}
%
The heavy neutrino mass spectrum is given by:
\begin{equation}
 	M_{1} = M_{11}\,\;\;\;\;\;\;M_{2}\cong M_{23} -\frac{A}{M_{23}}\,,
\;\;\;\;\;\;M_{3}\cong M_{23} + \frac{A}{M_{23}}\,,
 \end{equation}
%
with $\rho_3 = -\rho_2 = 1$. 
Note that $M_3 - M_2 = 2 A/M_{23} = m_3 - m_2$
and therefore, as in the preceding case,  
the splitting between $M_3$ and $M_2$
is exceedingly small and unobservable in practice.
The corrections to the matrix $V$ which diagonalises 
$M_N$ are of the order of $A/M^2_{23}$ and are negligible.
The elements of the matrix $R$, which parametrises the 
mixing between the light and the heavy neutrinos, have the form : 
$R^{*}_{\ell 1}=0$, $R^{*}_{\ell 2}=m^{D}_{\ell 3}/M_{23}$ and 
$R^{*}_{\ell 3}= m^{D}_{\ell 2}/M_{23}$, for $\ell= e,\mu,\tau$.
For the Majorana mass matrix for the LH flavour neutrinos 
we get an expression similar to the one in eq. (\ref{mnu2nu}):
$(m_{\nu})_{ll'} \cong -( m^{D}_{l2} m^{D}_{l'3} + 
m^{D}_{l3} m^{D}_{l'2})/M_{23}$.
If we assume that $|(m_{\nu})_{ll'}| \ltap 1$ eV
and 
and that $M_{23}\approx 100$ GeV, we obtain the following
constraint on the elements of the Dirac mass term:
\begin{equation}
 	|m^{D}_{\ell 2} m^{D}_{\ell^{\prime} 3}|
\ltap 10^{-7} {\rm GeV}^{2}\,\left(\frac{M_{23}}{100\,{\rm GeV}} \right)\,,~~
l,l'=e,\mu,\tau\,.
 \end{equation}
%
There are two distinct possibilities.\\
{\bf Democratic case:} $|m^{D}_{\ell 2}|$ and $|m^{D}_{\ell 3}|$ 
are of the same order. We have:
\begin{equation}
  |m^{D}_{\ell 2}|\approx |m^{D}_{\ell 3}|\ltap 3\times 10^{-4}\,{\rm GeV}\,.
\end{equation}
%
For $M_{23}\approx 100$ GeV, this case corresponds 
to exceedingly small couplings of the heavy Majorana 
neutrinos $N_{2,3}$ to the 
$W^\pm$ and $Z^0$: $|R_{\ell 2,3}|\ltap 3\times 10^{-6}$.
As a consequence, $N_{2,3}$ will be unobservable at LHC.\\ 
{\bf Hierarchical case:} suppose that $|m^{D}_{\ell 2}| \ll |m^{D}_{\ell 3}|$.
Consider, for instance the possibility:
\begin{equation}
m^{D}_{\ell 2}\approx 5\times 10^{-9}\,{\rm GeV}\,,\;\;\;\;\;\; m^{D}_{\ell 3}\approx 1\,{\rm GeV}\,.
 \end{equation} 
%
 This choice allows to 
have relatively large $|R_{\ell 2}|$, 
$|R_{\ell 2}|\approx 10^{-2}$ for $M_{23}\approx 100$ GeV, 
and thus relatively large $|(RV)_{\ell 2}|$ and  
$|(RV)_{\ell 3}|$. Thus, in principle, $N_2$ and $N_3$ can be 
produced with sufficiently large rates at, e.g. LHC, 
which might allow to observe them.
However, it would be hardly possible to obtain 
experimental evidences for their Majorana nature.
Indeed, one has $(RV)_{\ell 2} = (RV)_{\ell 3} = 
m^{D}_{\ell 3}/(\sqrt{2}M_{23})$. 
Therefore $N_2$ and $N_3$ couple 
to a given charged lepton $l$ (neutrino $\nu_l$) 
in the weak charged (neutral) current always in the 
combination $(N_2 + N_3)/\sqrt{2} = N_{PD}$. 
As a consequence, the magnitude of 
all physical effects 
associated with the Majorana nature of 
$N_2$ and $N_3$ will be determined by 
the mass difference $M_3 - M_2 = m_3 - m_2 \ltap 1$ eV,
which renders these effects unobservable 
in the experiments investigating the properties 
of the heavy neutrinos $N_{2,3}$. 

\paragraph{No Symmetry but $m_{\nu} = 0$ at Leading Order.} 
This case has been analised in detail recently in \cite{Amit10}, 
where also the general conditions for having 
$M_D\,M^{-1}_N\,M^T_D = 0$ have been derived.
We will consider one simple realisation 
of the indicated possibility. 
Namely, let us assume that $m^D_{l 1},m^D_{l 2} = 0$, but  
$m^D_{l 3} \neq 0$, $l=e,\mu, \tau$, and that the matrix $M_N$ 
has the form
\begin{eqnarray}
M_N = \left(
     \begin{array}{ccc} 
   M_{11} & 0 & 0 \\
     0   & 0 & M_{23}\\
     0   & M_{23} & M_{33}
\end{array}\right)\,.
\label{MN3}
\end{eqnarray}
%
In this case we have
$M_D\,M^{-1}_N\,M^T_D = 0$, and thus 
to leading order $m_{\nu} = 0$. 
Given the assumed simple structure 
of $M_D$ and $M_N$, the heavy Majorana 
neutrino $N_1$ having a mass $M_{11}$ decouples 
from the rest of the neutrino system, while $N_{2,3}$ 
couple via $W^{\pm}$  and $Z^0$ to the 
Standard Model particles.
For $M_{33} \neq 0$, 
there does not exist a conserved 
lepton charge and therefore higher order 
(one or two loop) contributions (see, e.g. \cite{Amit10}) 
lead to $m_{\nu} \neq 0$. Thus, this scheme does not 
belong to the class of see-saw scenarios (type I or inverse) 
which are the main subject of this study. 
Nevertheless, it is instructive to consider 
on this simple example the constraints
that have to be satisfied, which are specifically 
related to the presence of the heavy Majorana neutrinos.
It should be clear from the preceding discussion 
that all $|\Delta L| = 2$ Majorana type effects
should vanish in the limit of  $M_{33} = 0$. 

   Since $m_{\nu} = 0$ to leading order,
the constraint given in eq. (\ref{VR1}) 
is not applicable and the couplings of the 
two heavy Majorana neutrinos $N_{2,3}$ to the 
$W^\pm$ and $Z^0$ bosons can be relatively large. 
This in turn could lead to sufficiently large 
$N_{2,3}$ production  rates at LHC 
to make the observation of 
the two heavy Majorana neutrinos 
possible. We will show that 
the effects associated with the Majorana 
nature of $N_{2,3}$ are 
always proportional to the difference of the 
masses of $N_3$ and $N_2$, i.e. to $M_{33}$.
This is not surprising since in the limit 
of $M_{33} = 0$, there is a conserved lepton 
charge and all observable 
effects related to the Majorana 
nature of $N_{2,3}$ disappear.

  We will assume for simplicity in what follows 
that $m^D_{l 3}$,  $M_{23}$ and  $M_{33}$ 
are real and that $M_{23},M_{33} > 0$.
The diagonalisation of the neutrino mass 
Lagrangian (which includes the 
Dirac and Majorana mass matrices 
$M_D$ and $M_N$) shows that there are 
three massless mass-eigenstates $n_{1,2,3}$ and 
three massive Majorana mass-eigenstates 
$N_{1,2,3}$ with masses $M_1 = M_{11}$,
\begin{equation}
M'_{2,3} = \frac{1}{2}\, \left[ \sqrt{M^2_{33} + 4(M^2_{23} 
+ (m^D_{e 3})^2+ (m^D_{\mu 3})^2 +(m^D_{\tau 3})^2)} \mp M_{33} \right ]\,, 
\label{M1M2}
\end{equation}
satisfying the Majorana conditions:
$C \overline{N_{k}}^T =\rho_k N_k$, 
$k=2,3$, where $\rho_2 = -1$ and 
$\rho_3 = +1$.  In this case 
the angle $\theta$ of the 
$2\times 2$ orthogonal 
sub-matrix of the 
$3\times 3$ matrix ${\bf V}(\theta)$ 
which diagonalises $M_N$
is different, in general, from $\pi/4$:
we have $\cos^2\theta = M_3/(M_3 + M_2)$,
$\sin^2\theta = M_2/(M_3 + M_2)$,
where $M_{2,3} = (\sqrt{M^2_{33} + 4M^2_{23}} \mp M_{33})/2$
coincide up to the sign of $M_{2}$
with the 2nd and 3rd eigenvalues of the matrix $M_N$. 
We have $M_{3} - M_{2} = M'_{3} - M'_{2} = M_{33}$.
It is not difficult to find the matrices 
$R = M_D\,M^{-1}_N$ and 
$\eta = - R^*R^{\dagger}/2$: 
$\eta_{ll'} = - m^D_{l3}m^D_{l'3}/(2M^2_{23})$.
The existing limits on $|\eta_{l'l}|$ 
imply $|(m^D_{e3}/M_{23})^2| \ltap 8\times 10^{-3}$,
$|(m^D_{\mu 3}/M_{23})^2| \ltap 2.4\times 10^{-4}$.
Given these limits we have to a good approximation
$M'_{2,3} \cong M_{2,3}$.
The matrix $RV$ is given by:
\begin{eqnarray}
RV & = &\frac{1}{M_{23}}\left(\begin{array}{ccc}
 0 & m^D_{e3}\cos\theta  & m^D_{e3} \sin\theta\\
 0 &  m^D_{\mu 3} \cos\theta &  m^D_{\mu 3} \sin\theta\\
  0 &  m^D_{\tau 3} \cos\theta &  m^D_{\tau 3} \sin\theta
\end{array}\right)\;.
\label{RV2nu}
\end{eqnarray}
%

  For $M_{33} \ll 2M_{23}$, we recover the scheme with a 
heavy pseudo-Dirac neutrino: $N_{2,3}$ have different but 
close masses, $2(M'_3 - M'_2)/(M'_3 + M'_2) \cong 
M_{33}/\sqrt{M^2_{23} + (m^D_{e 2})^2+ (m^D_{\mu 2})^2} \ll 1$, 
$\theta \cong \pi/4$, and $N_{2,3}$ couple to the $W^{\pm}$ and 
$Z^0$ bosons only in the combination 
$N_{PD L} \cong (N_{2L} + N_{3L})/\sqrt{2}$.

 Consider the contribution to the $\betabeta$-decay 
effective Majorana mass due to the exchange of the 
two heavy Majorana neutrinos:
\begin{equation}
|(m_{\nu})_{ee}| \cong 
\left |\frac{(m^D_{e3})^2}{M^2_{23}}\,\left (M_3\,F(A,M_3)\sin^2\theta - 
 M_2\,F(A,M_2)\cos^2\theta \right) \right |\,.
\label{mee2}
\end{equation}
%
The function $F(A,M_k)$ to a very good approximation 
can be represented as (see, e.g. \cite{JV83,HaxStev84,Hirsch01,Blen10})
$F(A,M_k) \cong (M_a/M_k)^2 f(A,M_k)$, where 
$M_a \cong 0.9$ GeV and $f(A,M_k)$  
exhibits a weak dependence on $A$ and very weak 
dependence on $M_k$.
For $M_k \sim (100 - 1000)$ GeV of interest, 
the dependence of  $f(A,M_k)$ on 
$M_k$ is so weak \cite{Blen10} that  
can be safely neglected:
$f(A,M_2) \cong f(A,M_3) \equiv f(A)$.
Taking this into account we get:
\begin{eqnarray}
|(m_{\nu})_{ee}| &\cong& \left | \frac{(m^D_{e3})^2}{M^2_{23}}\,
\frac{M_a^2}{M_2\,M_3}\, \frac{M^2_3f(A,M_{2}) - 
M^2_2f(A,M_{3})}{M_3+M_2}\right | \\[0.30cm]
& \cong &
\left | \frac{(m^D_{e3})^2}{M^2_{23}}\,
f(A)\, M_a^2\, \frac{M_3-M_2}{M_2\,M_3}\right |
\cong \left |f(A)\,\frac{(m^D_{e3})^2}{M^2_{23}}\,
\frac{M^2_a}{M^2_{23}}\,M_{33} \right |\,, 
\label{mee3}
\end{eqnarray}
%
where we expressed the $\cos^2\theta$ and
$\sin^2\theta$ in terms of $M_{2,3}$.
For $M_{2,3}$ in the range of interest,
$M_k \sim (100 - 1000)$ GeV, and for, e.g. $^{76}$Ge,  
$^{82}$Se, $^{130}$Te and $^{136}$Xe, the function
$f(A)$ takes the following values 
\cite{Blen10} (see also \cite{HaxStev84})
$f(A)\cong$ 0.079, 0.073, 0.085 and 0.068, 
respectively; $f(A)$ has a somewhat smaller 
value of $^{48}$Ca: $f(^{48}Ca) \cong 0.033$.   
Given $M_a$ and $f(A)$,
the existing limits on $|(m_{\nu})_{ee}|$ imply a 
constraint on $((m^D_{e3})^2/M^2_{23})\,M_aM_{33}/M^2_{23}$.
If  $(m^D_{e3})^2/M^2_{23}$ will be determined from 
an independent measurement, the constraint on  $|(m_{\nu})_{ee}|$ 
will lead to a constraint on $M_{33}/M_{23}$. 
Taking, e.g. $|(m_{\nu})_{ee}| \ltap 1$ eV, $f(A) = 0.078$ (corresponding 
to $^{76}$Ge) and the maximal value of 
$(m^D_{e3})^2/M^2_{23}$ allowed by the data and quoted earlier, 
$8\times 10^{-3}$, one finds: 
$M_{33}\ltap 1.8 \times 10^{-5}\, M_{23} (M_{23}/M_a)$.
For $M_{23}=100$ GeV this implies  
$M_{33}\ltap 2\times 10^{-3} M_{23} 
\cong 0.2~{\rm GeV} \ll M_{23}$.
Such a small $N_2-N_3$ mass difference would render the 
Majorana-type effects associated with $N_{2,3}$
hardly observable. If, however, 
$|(m^D_{e3}/M_{23})^2| \ltap 1.6\times 10^{-6}$,
we get $M_{33}\ltap M_{23}$.

 Consider next the process of same sign di-muon 
production in $p-p$ collisions, assuming that 
one of the muons, say $\mu^-$, is produced together 
with real or virtual $N_{2,3}$ in the decay of 
a virtual $W^-$, while the second $\mu^-$ 
originates from  the decay $N_{2,3} \rightarrow W^+ + \mu^-$,
with virtual or real $W^+$ which decays further
into, e.g. two hadronic jets. 
The analysis is very similar to the one 
preceding eq. (\ref{ppmumu12}) - 
one has to replace $N_{1(2)}$ with 
$N_{2(3)}$, $M_{12}$ with $M_{23}$,
$M_{1(2)}$ with $M_{2(3)}$ and 
$\Gamma_{1(2)}$ with $\Gamma_{2(3)}$. 
The relevant part of the amplitude of the 
process under discussion can be obtained
from eq. (\ref{ppmumu12}) by 
replacing $m^D_{\mu 2}$ with $m^D_{\mu 3}$ 
and making the changes indicated above: 
\begin{equation}
P_{2,3} = \frac{(m^D_{\mu 3})^2}{M^2_{23}}\left [
 \frac{s^2 M_3}{p^2 - M^2_3 + i\Gamma_3 M_3}\, 
-\, \frac{c^2 M_2}{p^2 - M^2_2 + i\Gamma_2 M_2} \right ]\,,
\label{ppmumu}
\end{equation}
%
where now $p$ is the four momentum of the 
real (or virtual) $N_{2,3}$,
$c^2 = \cos^2\theta = M_3/(M_3 + M_2)$ and
$s^2 = \sin^2\theta = M_2/(M_3 + M_2)$.
Also in this case $p^2$ is equal to the square of the 
invariant mass of the two jets and the second muon
in the final state of the process 
(see, e.g. \cite{HanPasc09}).
We note that even when $N_{2}$ or $N_3$ is on mass shell, i.e. 
$p^2 = M^2_{2}$ or $p^2 = M^2_{3}$,
and one of the two terms in eq. (\ref{ppmumu})
dominates, the contribution of the 
second term (i.e. of the virtual $N_3$ or $N_2$)
is always present in the amplitude. 
The expression in eq. (\ref{ppmumu}) 
can be cast in the form:
\begin{equation}
P_{2,3} = \frac{(m^D_{\mu 3})^2}{M^2_{23}}\, \frac{M_2M_3}{M_3 + M_2}\,
\frac{M^2_3 - M^2_2 - i( \Gamma_3 M_3 - \Gamma_2 M_2)}
{(p^2 - M^2_3 + i\Gamma_3 M_3)(p^2 - M^2_2 + i\Gamma_2 M_2)}\,.
\label{ppmumu2}
\end{equation}
%
Taking into account that $M_2M_3 = 4M^2_{23}$, 
$(M_3 - M_2) = M_{33}$, and that  
$\Gamma_{2(3)} \propto G_F\, M^3_{2(3)}$ \cite{HanPasc09},
it is possible to show that $P_{2,3}$ 
vanishes in the limit of $M_3 = M_2$:
$P_{2,3}\propto (M_3 - M_2) = M_{33}$.
Thus, if $M_{33} \ll M_{23}$, the amplitude 
of the process $p+p\rightarrow \mu^- + \mu^- + 2~{\rm jets} + X$,
generated by the production and decay of 
real or virtual $N_{2,3}$, will be strongly 
suppressed.

\paragraph{The Extreme Fine-tuning Case.} It is well known that
the see-saw mechanism is underconstrained, namely there
is an infinite set of Dirac neutrino mass matrices leading 
to the observed neutrino parameters. The most general
Dirac neutrino mass matrix that satisfies $m_\nu=-M_D M_N^{-1} M^T_D$,
with $m_\nu\simeq U_{\PMNS}^* \hat m U_{\PMNS}^\dagger$ and 
$M_N\simeq V^* \hat M V^\dagger$, can be parametrized as~\cite{Casas:2001sr}:
\begin{equation}
M_D= i U_{\PMNS}^* \sqrt{\hat m} \Omega \sqrt{\hat M} V^\dagger\;,
\label{mDfromOmega}
\end{equation}
%
where $\Omega$ is an arbitrary
complex orthogonal matrix. Hence, by choosing conveniently the matrix
$\Omega$, it is always possible to find a Dirac neutrino mass matrix with
at least one large eigenvalue leading to the observed neutrino parameters, 
while keeping the right-handed neutrino masses in the range $ (100-1000)$ GeV. 
However, as we will show below, this possibility requires in general a 
huge tuning of parameters.

Let us consider for simplicity the two-generation case. Then,
the $\Omega$-matrix can be decomposed in:
\begin{equation}
\Omega=\left(\begin{array}{cc} 
   \cos\hat\theta & \sin\hat\theta \\
    -\, \sin\hat\theta   &  \cos\hat\theta
  \end{array}\right)=
\frac{e^{i\hat\theta}}{2}\left(\begin{array}{cc} 
  1 & -i \\
  i  & 1 
  \end{array}\right) +
\frac{e^{-i\hat\theta}}{2}\left(\begin{array}{cc} 
  1 & i \\
  -i  & 1 
  \end{array}\right) \equiv \Omega_+ +\Omega_-\;,
\end{equation}
where $\hat \theta=\omega -i \xi $ is a complex parameter.
Accordingly, the Dirac neutrino mass matrix can be decomposed
as $M_D=M_{D+}+M_{D-}$, in a self-explanatory notation.

Taking for definiteness $\xi>0$, it follows that $M_{D+}$ ($M_{D-}$) 
grows (decreases) exponentially with  $\xi$. Therefore, for sufficiently
large  $\xi$ it is possible to compensate the huge suppression
in eq. (\ref{mDfromOmega}) from the tiny observed neutrino masses 
and the light right-handed neutrino masses. Note however that $M_{D-}$
cannot be neglected, even though it is exponentially suppressed 
compared to $M_{D+}$, since the naive approximation 
$M_D\simeq M_{D+}$ leads to $m_\nu= 0$, due to $\Omega_+ \Omega_+^T=0$.
Therefore, reproducing the correct neutrino parameters requires
a large amount of tuning, concretely
\begin{equation}
\frac{(M_{D-})_{ij}}{(M_{D+})_{ij}}\sim e^{-2\xi} \sim
\frac{m_i M_j}{(M_{D})^2_{ij}}\;.
\label{fine-tuning}
\end{equation}
For instance, demanding $(M_{D})_{ij}\sim{\cal O}(1\,{\rm GeV})$
and $M_j \sim{\cal O}(100\,{\rm GeV})$ requires a tuning of one
part in $10^9$ in order to produce a neutrino mass 
$m_i \sim{\cal O}(10^{-2}\,{\rm eV})$.
The fine-tuning problem of this scenario is exacerbated by the
presence of radiative corrections to the see-saw parameters
which usually spoil the tuning, unless the radiative corrections
to the different parameters are highly correlated in such a way that
the tuning is preserved. This possibility is extremely unnatural
unless originated by an underlying approximate symmetry, such the
one proposed before eq.(\ref{MN2}). 

Furthermore, such light right-handed neutrinos with such large 
couplings can induce a rate for $(\beta\beta)_{0\nu}$-decay 
in conflict with the experimental constraints. 
The contribution from the right-handed neutrinos to the 
$(\beta\beta)_{0\nu}$-decay is approximated by:
\begin{equation}
|(m_{\nu})_{ee}| \cong 
\left |\sum_k\, F(A,M_k)\, (RV)^2_{e k}\,M_k \right |\,,
\end{equation}
where in this case 
\begin{equation}
RV=-i U_{PMNS} \sqrt{\hat m} \Omega^* \sqrt{\hat M^{-1}}\,.
\end{equation}
Using as before that $F(A,M_k) \cong (M_a/M_k)^2 f(A,M_k)$, where 
$M_a \cong 0.9$ GeV and $f(A,M_k)$ has a weak dependence with $M_k$,
we finally obtain:
\begin{equation}
|(m_{\nu})_{ee}| \cong 
\left |\sum_k\, (U_{\PMNS} \sqrt{\hat m} \Omega^*)^2_{ek} 
\frac{M^2_a}{M_k^2} f(A,M_k)\right|\,.
\end{equation}

The dominant contribution to this expression is given by 
the exponentially-enhanced $\Omega_+$ matrix, yielding:
\begin{eqnarray}
|(m_{\nu})_{ee}| &\cong &
\left |\frac{e^{2\xi}}{4}\frac{M_2^2-M_1^2}{M_1^2 M_2^2} 
[\sqrt{m_1}(U_{\PMNS})_{11}-i\sqrt{m_2}(U_{\PMNS})_{12}]^2 
f(A) M_a^2  \right| \nonumber \\
&\cong&
\left|\frac{1}{4}\frac{M_{D}^2}{M_2^2}\frac{M_a^2}{M_2^2}f(A)(M_2-M_1)  \right|
\cong 10^{-10} (M_2-M_1)\,,
\end{eqnarray}
for $M_{D}\sim{\cal O}(1\,{\rm GeV})$ 
and $M_2 \sim{\cal O}(100\,{\rm GeV})$. Therefore, the non-observation 
of the $(\beta\beta)_{0\nu}$-decay requires in this scenario a 
degeneracy in the right-handed neutrino masses of at least one per cent.
As discussed above, the cross section for same sign di-muon production
in $p-p$ collisions is proportional to the mass difference of the 
right-handed neutrinos. Thus, even in this extremely fined-tuned scenario, 
the Majorana nature of the right-handed neutrinos will be difficult to probe
at colliders.

%
\section{Multiple Mass Scale See-Saw Scenarios}
%

 We will consider next versions of the see-saw 
scenario, in which we allow couplings of 
the RH neutrinos with other SM singlets
that are therefore involved in the 
mechanism of generation of light neutrino masses. 
This implies the presence of more than two 
mass scales in the latter. 

%
\subsection{Scenario 1}
%

Consider the following mass Lagrangian:
\begin{equation}
 \mathcal{L}_{\mu}\;=
-\;\overline{\nu_{\ell L}} (m_{D})_{\ell a}\nu_{aR} \;
- \; \overline{S_{\beta L}}(M_{R})_{\beta a}\nu_{aR}\;
-\;\half \overline{S}_{\beta L}(\mu)_{\beta \beta'}S^{C}_{\beta' R}\;+\;{\rm h.c.} 
\label{LInvSeeSawS1}
\end{equation}
%
where $S^{C}_{\beta'R} \equiv C \overline{S_{\beta' L}}^T$. 
We have introduced an arbitrary fixed numbers 
of RH neutrinos $\nu_{aR}$ 
and left-handed SM gauge singlets $S_{\beta L}$.
We comment on their numbers below.

  In what follows we assume that the scale of 
$M_R$ is much bigger than the scales of $m_D$ and $\mu$. 
If we assign one unit of the total lepton charge $L$ to
$\nu_{\ell L}$, $\nu_{aR}$ and $S_{\beta L}$, the terms 
involving the mass matrices $m_D$ and $M_R$ conserve $L$, 
while the term with $\mu\neq 0$ changes $L$ by 2 units.
Thus, the $\mu$-term breaks explicitly the $U(1)$ symmetry 
associated with the lepton charge conservation. 
In the limit of  $\mu = 0$, there is a conserved 
lepton charge and the particles with  
definite mass are either massless or are massive 
Dirac fermions. Given the number of the LH flavour neutrino 
fields $\nu_{l L}$, $n(\nu_L)$, the numbers of massless and 
massive Dirac states depends \cite{LeungSTP83} 
(see also \cite{BiPet87})
on the number of RH neutrino fields 
$\nu_{a R}$, $n(\nu_R)$, and on
the number of LH singlets $S_{\beta L}$, $n(S_L)$.
If, for instance, we have $n(\nu_{L}) = n(\nu_R) = n(S_L) =1$, 
there is one massless and one massive Dirac neutrinos.
In the case of  $n(\nu_{L}) = n(\nu_R) = 3$ and , e.g.  $n(S_L) =1$, 
we will have 3 massive Dirac  states 
and one massless neutrino. 
In the general case the numbers 
of massive Dirac and massless states are given by 
\cite{LeungSTP83} $N_D = {\rm min}(n_{L} + n(S_L),n(\nu_R))$ 
and  $N_{0} = |n_{L} + n(S_L) - n(\nu_R)|$, respectively.
Thus, if $n(S_L) = n(\nu_R)$, the number of massless states 
coincides with the number of the LH flavour neutrinos.
This is the case we will be interested in what follows.
In this case the three massless states 
acquire nonzero Majorana masses when $\mu \neq 0$.
At the same time each massive Dirac neutrino is split into
two Majorana neutrinos having different but very close masses. 

 In view of the above one can expect that 
the light neutrino Majorana mass matrix 
depends linearly on $\mu$ in such a way that 
in the limit $\mu\to 0$, the lepton charge conservation 
symmetry is restored and the three LH flavour 
neutrinos become massless. The heavy neutrino sector 
is given by the mixing of the fields  
$\nu_{aR}$ and $S_{\beta L}$, with relatively small mass 
splittings. This scenario is the well known 
inverse see-saw model \cite{Mohapatra:1986bd,LWDW83}.

  Formally, we can derive the expressions of the 
light neutrino mass matrix $m_{\nu}$ and the corresponding 
non unitarity mixing parameters  $\eta$
from expressions (\ref{R}), (\ref{m2}) and (\ref{eta}), 
by replacing the matrices $M_D$ and $M_R$ with
\footnote{In principle, one can add a non zero 
$k\times k$ block in the $11$ entry of the block
Majorana mass matrix $M_{N}$ in eq. (\ref{MNS1}), which acts 
as a small perturbation 
that breaks lepton number explicitly. However, it can be proven 
that such term does not enter
in the expression of the light neutrino mass matrix 
(see, e.g. \cite{Dudas:2002ry} for an explicit model) 
and, therefore, we do not consider this case. }
\begin{eqnarray}
 M_D \;\equiv\; \left(
      \begin{array}{cc}
       m_D & \N 
      \end{array}\right)\;\;\;\;\;\;
 M_N\equiv\left(
     \begin{array}{cc} 
     \N & M_R^T\\
     M_R & \mu                
     \end{array}\right)\label{MNS1}
\end{eqnarray}
%
We assume further that $M_R\gg m_D > \mu$. We note that the 
parameters in the $\mu$ term in the lagrangian (\ref{LInvSeeSawS1}) 
 can be arbitrarily small because, as we have already 
noticed before, this is the term in the Lagrangian that 
breaks explicitly the lepton number.
The actual scale of the $\mu$ term depends on the model considered, 
which at low energy is reduced to an effective field 
theory described by the Lagrangian (\ref{LInvSeeSawS1}) . 
This mass scale can, indirectly, affect the non-unitarity 
effects in the neutrino mixing as well as the couplings 
of the heavy singlet Majorana fields to the EW gauge bosons, 
due to the interplay with the other scales in the theory, 
namely, $m_D$ and $m_R$ in the see-saw mass 
formula (see eq. (\ref{mnuS1})). 

The full mass matrix corresponding to 
eq. (\ref{LInvSeeSawS1}) takes the form:
\begin{eqnarray}
\mathcal{M} \equiv\left(
     \begin{array}{ccc} 
   \N & m_D & \N \\
   m_D^T & \N &  M_R^T\\
   \N &  M_R & \mu                
     \end{array}\right)\,.
\end{eqnarray}
%
From eq. (\ref{R}) we obtain:
\begin{eqnarray}  
 R^\dagger  & = & \left(
    \begin{array}{c}
       - M_R^{-1} \mu \left(M_R^{-1}\right)^T m_D^T \\
        \left(M_R^{-1}\right)^T m_D^T
    \end{array}\right) \,.
\end{eqnarray}
%
Consequently, light neutrino Majorana mass matrix is given by:
\begin{eqnarray}
 m_{\nu} =  U^* \hat{m} U^{\dagger} & \simeq & - R^{*} M_{N} R^{\dagger} \;=\; m_{D} M_{R}^{-1} \mu \left( M_{R}^{-1}\right)^{T}m_{D}^{T} \label{mnuS1}\,.
\end{eqnarray}   
%
The expression for the non unitary correction  matrix $\eta$ includes two terms:
\begin{eqnarray}
 \eta & = & -\half m_{D}^{*}  \left(M_{R}^{-1} \right)^{*} \mu^{*} \left(M_{R}^{-1} \right)^{\dagger}  \left(M_{R}^{-1} \right) \mu \left(M_{R}^{-1} \right)^{T} m_{D}^{T} 
 \;-\; \half m_{D}^{*}\left(M_{R}^{-1} \right)^{\dagger} \left(M_{R}^{-1} \right) m_{D}^{T}\label{etaS1}\,.
\end{eqnarray}
According to the hierarchy of the mass scales that enter in the theory, the second term is the dominant one and it does not depend on the
LNV parameters in  $\mu$. Therefore, it is possible to have an observable deviation from unitarity of the PMNS neutrino
mixing matrix, without interfering with the tight constraints on the neutrino mass scale, $m_{\nu}\lesssim 1$ eV, which is proportional
to the  ``small'' Majorana mass matrix $\mu$. If we take large non-unitarity effects $\eta \approx 10^{-4}$, 
and right-handed fields at the scale $M_{R}\approx 1$ TeV,
the lepton number breaking parameters in the lagrangian are given at the scale:
\begin{equation}
 \mu\;\approx\; \frac{m_{\nu}}{\eta}\;\approx\; 10\; {\rm keV}\,.
\end{equation}
%
This estimate shows that if the scenario 
can be tested in neutrino experiments and experiments studying
LFV processes (see $e.g.$ \cite{Abada:2007ux}), 
the generation of the $\mu$ term cannot be
associated with the EW symmetry breaking. Indeed, assuming new physics 
at the scale $\Lambda_{NP}\approx1$ TeV, a coupling of the form 
$(1/\Lambda_{NP})H^{\dagger}H\overline{S_{L}}S^{C}_{R}$ 
(see, e.g. \cite{Bonnet:2009ej}) implies:
\begin{equation}
 \mu\;\approx\; \frac{v^{2}}{\Lambda_{NP}}\;\approx\;10\; {\rm GeV}\,. 
\end{equation}
%
It follows from the see-saw mass formula 
that in this case the non-unitarity effects and the couplings 
of the heavy neutral fermions to SM particles
are respectively $\eta\approx10^{-11}$ and 
$RV \approx10^{-6}$, which are 
too small to produce measurable 
effects in the ongoing and the planned future 
experiments.

Apart from the possibility of observing 
sizable deviations from unitarity of the 
light neutrino mixing in the forthcoming
experiments, 
in this scenario the production of the 
Majorana SM singlets at colliders might not be suppressed.
However, all lepton charge violating (LCV) processes 
involving the heavy singlets, which are associated with their 
Majorana nature, are strongly suppressed, which renders 
them unobservable in the current and currently planned 
future experiments. Therefore, the heavy Majorana singlets, 
even if produced with sufficiently large rates to 
be observable, will behave like heavy Dirac neutral 
singlets to a relatively high level of precision. 

  In order to illustrate this point, 
we consider for simplicity the 
case $n(\nu_{L})=2$ and $n(\nu_{R})=n(S_{L})=1$.  We assume
all parameters in the theory to be real with positive 
$M_{R}$ and $\mu$. The Dirac mass term is 
simply $m_{D}^{T}=(m_{e1}^{D}\;\;m_{\mu1}^{D})$.
Therefore, the non unitary part of 
the PMNS neutrino mixing matrix reads:
\begin{equation}
 \eta\;\cong\;\frac{1}{M_{R}^{2}}
 \left(
      \begin{array}{cc}
       (m_{e1}^{D})^{2} & m_{e1}^{D} m_{\mu1}^{D}  \\
       m_{e1}^{D} m_{\mu1}^{D} & (m_{\mu1}^{D})^{2}   
      \end{array}\right)\,.
\end{equation}
%
In the framework considered, the particle content of 
the theory is given by one massless neutrino, 
a light Majorana neutrino with mass 
$m_{\nu}=\mu(m_{D}/M_{R})^{2}$ and two
heavy Majorana neutrinos $N_{1,2}$ having 
different but close masses $M_1 \neq M_2$ of the order of 
$M_{1,2}\cong M_{R}\approx 100\div1000$ GeV
and a mass splitting $|M_{1}-M_{2}|\approx\mu$.
The heavy Majorana neutrino fields satisfy 
the Majorana conditions:
$C\overline{N_{1,2}}^{T}\equiv \rho_{1,2}N_{1,2}$, 
with $\rho_{1}=-1$ and $\rho_{2}=1$. 
In the limit $\mu=0$, the lepton charge symmetry 
is restored and the spectrum consists of two massless LH neutrinos 
and a Dirac heavy neutrino $N_{D}\equiv(-N_{1}+N_{2})/\sqrt{2}$,
with $\nu_{R}\equiv N_{DR}$. Therefore, in the LCV 
regime ($\mu\neq0$) we have a heavy pseudo-Dirac neutrino 
field $N_{PD}\equiv(-N_{1}+N_{2})/\sqrt{2}$ which is coupled 
to the EW gauge bosons via the neutrino mixing. 
Indeed, the heavy LH components of the two heavy Majorana fields
$N_{1}$ and $N_{2}$ have the following couplings to the $W^{\pm}$ and $Z^{0}$ 
bosons:
\begin{equation}  
RV \;=\;\frac{1}{\sqrt{2}}
 \left(
      \begin{array}{cc}
      -\frac{\mu}{M_{R}} \frac{m^{D}_{e1}}{M_{R}}-  \frac{m^{D}_{e1}}{M_{R}}& -\frac{\mu}{M_{R}} \frac{m^{D}_{e1}}{M_{R}} +  \frac{m^{D}_{e1}}{M_{R}}  \\
       -\frac{\mu}{M_{R}} \frac{m^{D}_{\mu1}}{M_{R}} -  \frac{m^{D}_{\mu1}}{M_{R}} & -\frac{\mu}{M_{R}} \frac{m^{D}_{\mu1}}{M_{R}} +  \frac{m^{D}_{\mu1}}{M_{R}}   
      \end{array}\right)\,.
\end{equation}
%
Consequently, the CC Lagrangian (\ref{NCC}) can be cast 
in the form:
\begin{equation}
 \mathcal{L}_{CC}^{N_{PD}}= -\frac{g}{2\sqrt{2}}\left(\epsilon_{\ell} \,\overline{\ell}\gamma_{\alpha}(1-\gamma_{5})N_{PD}
 \,+\,\epsilon^{\prime}_{\ell}\overline{N_{PD}}\gamma_{\alpha}(1+\gamma_{5})\ell^{C} \right)W^{\alpha}\;+\;{\rm h.c.}\,,
\end{equation}
%
where $\ell=e,\mu$ and 
\begin{equation}
 \epsilon_{\ell}\;=\;\frac{m^{D}_{\ell1}}{M_{R}}\;\;\;\;\;\;\;\;\;\;\;\;\;\;\;\;\; \epsilon^{\prime}_{\ell}\;=\; \frac{\mu}{M_{R}}\frac{m^{D}_{\ell1}}{M_{R}}\,.
\end{equation}
%
Therefore, similarly to the previous scenario, 
large production rates of the heavy Majorana 
neutrinos $N_{1,2}$ are possible at colliders, but 
LCV decays (processes) associated with their Majorana nature
are strongly suppressed. Indeed, 
the suppression factor for the rate of the LCV decay 
$N_{PD} \to \ell^{+} W^{-}$ is given by: 
$|\epsilon^{\prime}_{\ell}|^{2}\cong m_{\nu}^{2}/(\eta M_{R}^{2})$, 
and for $\eta\approx10^{-4}$ and $M_{R}\approx 100$ GeV 
we have $|\epsilon^{\prime}_{\ell}|^{2}\approx 10^{-18}$.

%
\subsection{Scenario 2}
%

We consider now a variation of the previous scenario in which 
the source of the lepton number breaking parameter is a
(small) Dirac-type mass term between the heavy singlets. The 
Dirac and Majorana neutrino mass matrices 
of this model are the following:
\begin{eqnarray}
 M_D \;\equiv\; \left(
      \begin{array}{cc}
       m_D & \mu^{\prime} 
      \end{array}\right)\;\;\;\;\;\;
 M_N\equiv\left(
     \begin{array}{cc} 
     \N & M_R^T\\
     M_R & \N                
     \end{array}\right)
\end{eqnarray}
%
We assume also in this case a hierarchical mass pattern: 
$M_{R} \gg m_{D},\,\mu^{\prime}$. The neutrino mass matrix 
and the deviation of $U_{\PMNS}$ 
from unitarity in the scheme considered are given by:
\begin{eqnarray}
  U^* \hat{m} U^{\dagger} & \simeq & - R^{*} M_{N} R^{\dagger} \;=\; -m_{D} M_{R}^{-1} \mu^{\prime T} - \mu^{\prime} \left( M_{R}^{-1}\right)^{T} m_{D}^{T}\,, \\
  \eta & = & -\half m_{D}^{*} \left( M_{R}^{-1}\right)^{*} 
\left( M_{R}^{-1}\right)^{T} m_{D}^{T}  -\half \mu^{\prime *} 
\left( M_{R}^{-1}\right)^{\dagger} \left( M_{R}^{-1}\right) \mu^{\prime T}\,.
\end{eqnarray} 
%
In the case considered, a large mixing between the light  
and heavy singlet neutrinos 
corresponds to a much smaller 
lepton number breaking scale $\mu^{\prime}$, 
which is given roughly 
by $\mu^{\prime}\approx m_{\nu}/\sqrt{|\eta|}\approx 10$ eV.  
It is not difficult to prove that also 
in this case the lepton number non-conserving 
couplings of the heavy singlet neutrinos  to the
EW gauge bosons $W^{\pm}$ and $Z^0$
are exceedingly small which makes the $|\Delta L| = 2$ effects 
unobservable:
the couplings of interest are given 
approximately by $m_{\nu}/(\sqrt{|\eta|}M_R)$ 
and for, e.g. $\eta\approx 10^{-9}$ and $M_{R}\approx100$ GeV
we have $|m_{\nu}|/(\sqrt{|\eta|} M_{R})\approx 10^{-7}$.

\section{Avoiding the Constraints: Non-Singlet Heavy Neutrinos} 

The previous general argument shows that the requirement 
$|(m_{\nu})_{ll'}| \lesssim 1$ eV, $l,l'=e,\mu,\tau$, translates
into an extremely suppressed charged and neutral current interactions 
of the heavy Majorana fields $N_j$ with the Standard Model charged 
leptons and neutrinos, unless the heavy Majorana neutrinos form
a pseudo-Dirac pair. Therefore, if these interactions are the
only portal to the Standard Model, the Majorana nature of the heavy
neutrinos will not be detected in collider experiments: either the
production cross section is highly suppressed or the heavy neutrinos
behave to a high level of precision as Dirac fermions. 

  This result may not be valid if there exist  
additional TeV scale interaction terms in the Lagrangian between 
the heavy Majorana neutrinos and the Standard Model particles. 
If this is the case, the production cross section of heavy 
neutrinos will not necessarily be suppressed, while 
their charged and neutral current interactions 
with the Standard Model charged leptons and neutrinos can still 
be tiny.

  One possibility is the existence of 
an extra $U(1)$ local gauge symmetry, 
which is broken at the 
TeV scale and under
which the Standard Model particles 
and the heavy (RH) neutrinos are charged 
(see, e.g. \cite{SKhalil10}). 
In this case, the production cross section of two 
heavy Majorana neutrinos can be largely enhanced. 
At the same time, the heavy neutrinos can decay 
only into Standard Model 
particles and they can do it 
only through the tiny charged current and neutral current 
couplings. This implies that the heavy neutrinos will 
be relatively long-lived and thus will have 
a relatively large decay length
which in turn  will yield a characteristic 
displaced vertex in the detectors 
(see, e.g. \cite{Strum08}). 
More importantly, 
if the heavy neutrinos are true Majorana particles, 
their production and decay will lead to events with 
a pair of same-sign muons in the final state. 
The cross section can be large enough to allow the 
observation of this lepton number
violating process at colliders.

   A second example can be found in the TeV scale 
type III see-saw mechanism. In this case, 
the heavy states form an $SU(2)_L$ triplet of leptons,
$L^{\pm}$, $L^0$, with essentially the same mass, 
$L^{\pm}$ being somewhat heavier than $L^0$ 
(see, e.g. \cite{Strum08}).
A pair of these leptons, say $L^{+}$ and $L^0$,
can be produced in colliders via their gauge coupling to 
the $W^{\pm}$-boson.  The charged heavy lepton 
$L^{+}$ can decay into $\mu^+ + Z^0$.
The heavy neutral Majorana lepton $L^0$
has an interaction Lagrangian with the Standard 
Model charged leptons and neutrinos, which
is similar to that given in eqs. (\ref{nuCC}) and (\ref{nuNC}).
Thus, being a Majorana particle, $L^0$ can decay into 
$\mu^+ + W^{-}$, leading to same-sign dimuon (plus 4 jets) 
events with observable displaced vertices 
of the two muons in the detectors.
Detailed calculations have shown that for masses of 
$L^{+}$ and $L^0$ not exceeding 1000 GeV, the 
$\mu^+\mu^+$ + 4 jets events can have 
observable rates at LHC (see, e.g. \cite{Strum08}).

\section{Conclusions}

  In this article we have discussed the 
possibility to test the Majorana nature 
of the heavy 
Majorana neutrinos $N_j$
which are an integral part of the TeV scale
type I
and inverse see-saw  scenarios of neutrino 
mass generation.
In the indicated TeV scale see-saw scenarios 
the heavy Majorana
neutrinos typically have masses 
in the range of $M_j \sim (100 - 1000)$ GeV.
The fact that  $N_j$ are Majorana particles 
can be revealed by observation of processes 
with real or virtual $N_j$, in which the total 
lepton charge $L$ changes by two units, $|\Delta L| = 2$, 
like $p + p \rightarrow \mu^- + \mu^- + 2jets + X$ 
at LHC, etc.
We have shown that
the physical effects associated with 
the Majorana nature of these heavy 
neutrinos $N_j$, 
are so small that they
are unlikely to be observable
in the currently operating and 
future planned accelerator experiments 
(including LHC). This is a consequence
of the existence of very strong constraints 
on the parameters and couplings, responsible for the 
corresponding $|\Delta L| = 2$ processes in which 
$N_j$ are involved, and/or
on the couplings of $N_j$ 
to the weak $W^{\pm}$ 
and  $Z^0$ bosons. 
The constraints are related to the fact 
that the elements of the Majorana mass matrix 
of the left-handed flavour neutrinos $m_{\nu}$, 
generated by one of the indicated 
see-saw mechanisms, 
should satisfy, in general, 
$|(m_{\nu})_{ll'}| \lesssim 1$ eV, $l,l'=e,\mu,\tau$;
in the case of the  $(m_{\nu})_{ee}$ element, 
the upper limit follows from the 
experimental searches for neutrinoless 
double beta ($\betabeta$-) decay.
Even in the case of extreme fine tuning 
(at the level of one part in $10^{9}$ 
or $10^{10}$), in which the neutrino 
Yukawa couplings can be of order 1,
the upper limit on $|(m_{\nu})_{ee}|$
obtained in the $\betabeta$-decay 
experiments implies a strong constraint 
on the  $|\Delta L| = 2$ 
heavy Majorana neutrino mass 
splitting(s) (or masses), 
which makes it very difficult   
(if not impossible) to probe 
the Majorana nature of the 
heavy Majorana neutrinos 
in experiments at colliders.
The simple illustrative examples we have considered
suggest that
if the heavy Majorana neutrinos $N_j$
are observed and they are associated with the 
type I
or inverse see-saw mechanisms 
and no additional TeV scale  ``new physics'', 
they will behave
like Dirac fermions to a relatively 
high level of precision, 
being actually pseudo-Dirac particles.
The observation of effects proving 
the Majorana nature of $N_j$ would imply that 
these heavy neutrinos
have additional relatively strong
couplings to the 
Standard Model particles (as, e.g. in the 
TeV scale type III see-saw scenario), 
or that the light neutrino masses compatible with the 
observations are generated 
by a mechanism other than the see-saw 
(e.g., radiatively at one or two loop level) 
in which the heavy singlet
Majorana neutrinos $N_j$ 
are nevertheless involved. 

   The considerations presented in this article 
and the conclusions reached concern a rather large number 
of TeV scale see-saw models discussed in the literature 
(see, e.g. \cite{TeVsees,Bonnet:2009ej}).

\vspace{-0.4cm}
\section{Acknowledgments}
\vspace{-0.3cm}
%
We would like to thank S. Pascoli for discussions. 
S.T.P. whishes to thank 
M. Blennow, E. Fernandez-Martinez, 
J. Lopez-Pavon and J. Menendez 
for useful correspondence.
This work was supported in part by 
the INFN program on ``Astroparticle Physics'', 
by the World Premier International Research Center 
Initiative (WPI Initiative), MEXT, Japan (A.I. and S.T.P.), 
the EU-RTN Programme (Contract No.MRTN--CT-2006-035482) 
Flavianet (E.M.) and by the DFG cluster of excellence ``Origin 
and Structure of the Universe'' (A.I.).


\begin{thebibliography}{99}


\bibitem{cleveland98} B.T. Cleveland  {\it et al.}, Astrophys.\
J. \textbf{496}, 505 (1998).

\bibitem{fukuda96} Y. Fukuda {\it et al.} [Kamiokande Collaboration],
Phys.\ Rev.\ Lett. \textbf{77}, 1683 (1996).

\bibitem{abd09} J.N. Abdurashitov {\it et al.}, Phys.\ Rev.\ C \textbf{80},
015807 (2009).

\bibitem{anselmann92} P. Anselmann {\it et al.}, 
Phys.\ Lett.\ B \textbf{285} (1992) 376;
%
W. Hampel {\it et al.}, 
Phys.\ Lett.\ B \textbf{447}  (1999) 127;
%
M. Altmann {\it et al.}, Phys.\ Lett.\ 
B \textbf{616}  (2005) 174.


\bibitem{SKsolar02}
 S.~Fukuda {\it et al.} [Super-Kamiokande Collaboration],
Phys. Lett. {\bf B539} (2002) 179.


\bibitem{ahmad01} Q.R. Ahmad {\it et al.} [SNO Collaboration], 
Phys.\ Rev.\ Lett. \textbf{87} (2001) 071301 and
\textbf{89} (2002) 011301.

\bibitem{SKatm98}
Y.~Fukuda {\it et al.}  [Super-Kamiokande Collaboration],
Phys.\ Rev.\ Lett.\  {\bf 81} (1998) 1562.

\bibitem{SKdip04} Y. Ashie {\it et al.} [Super-Kamiokande
                 Collaboration], 
Phys. Rev. Lett. {\bf 93} (2004) 101801.

\bibitem{KL162}
K.~Eguchi {\it et al.}  [KamLAND Collaboration],
Phys.\ Rev.\ Lett.\  {\bf 90} (2003) 021802; 
%
T. Araki {\it et al.},
Phys.\ Rev.\ Lett. \textbf{94} (2005) 081801.

\bibitem{BOREX} C. Arpesella {\it et al.}, 
Phys. Lett. B {\bf 658} (2008) 101; 
Phys. Rev. Lett. {\bf 101} (2008) 091302.

\bibitem{ahn06} M.~H.~Ahn {\it et al.}  [K2K Collaboration],
Phys.\ Rev. \ D {\bf 74} (2006) 072003.


\bibitem{Michael06} D.G. Michael {\it et al.} [MINOS Collaboration], 
Phys.\ Rev.\ Lett. \textbf{97} (2006) 191801; 
P. Adamson {\it et al.},
Phys.\ Rev.\ Lett. \textbf{101} (2008) 131802.


\bibitem{BPont57} B. Pontecorvo, 
                  Zh. Eksp. Teor. Fiz. (JETP) {\bf 33} (1957) 549 
                and {\bf 34} (1958) 247.

\bibitem{MNS62} Z. Maki, M. Nakagawa and S. Sakata, 
Prog. Theor. Phys. {\bf 28} (1962) 870.

\bibitem{BPont67}
  B.~Pontecorvo,
  Sov.\ Phys.\ JETP {\bf 26} (1968) 984.
  [Zh.\ Eksp.\ Teor.\ Fiz.\  {\bf 53}, 1717 (1967)].

\bibitem{Znu}  
  C.~Amsler {\it et al.}  [Particle Data Group],
  Phys.\ Lett.\  B {\bf 667} (2008) 1.

\bibitem{seesaw}  P. Minkowski,
Phys.\ Lett.\  B {\bf  67} (1977) 421;
M. Gell-Mann, P. Ramond and R. Slansky, 
{\em Proceedings of the Supergravity Stony Brook Workshop}, 
New York 1979,  eds. P. Van Nieuwenhuizen and D. Freedman;
T. Yanagida,  
{\em Proceedinds of the Workshop on Unified Theories and Baryon Number in theUniverse},  Tsukuba, Japan 1979, ed.s A. Sawada and A. Sugamoto;
 R. N. Mohapatra and G. Senjanovic, Phys. Rev. Lett. {\bf 44} (1980) 912.
  
\bibitem{BiPet87} S.~M.~Bilenky and S.~T.~Petcov,
  Rev.\ Mod.\ Phys.\  {\bf 59} (1987) 671.


\bibitem{Foot:1988aq}
  R.~Foot, H.~Lew, X.~G.~He and G.~C.~Joshi,
  Z.\ Phys.\  C {\bf 44} (1989) 441.

\bibitem{Mohapatra:1986bd} 
  R.~N.~Mohapatra and J.~W.~F.~Valle,
  Phys.\ Rev.\  D {\bf 34} (1986) 1642.

\bibitem{TeVsees}  J.~Chakrabortty,
  arXiv:1003.3154; 
C.~Wei,
  arXiv:1003.1468;  
  F.~M.~L.~de Almeida {\it et al.}, 
  Phys.\ Rev.\  D {\bf 81} (2010) 053005;
  H.~Zhang and S.~Zhou,
  Phys.\ Lett.\  B {\bf 685} (2010) 297;
  P.~S.~B.~Dev and R.~N.~Mohapatra,
  Phys.\ Rev.\  D {\bf 81} (2010) 013001;
  Z.~z.~Xing and S.~Zhou,
  Phys.\ Lett.\  B {\bf 679} (2009) 249.


  
\bibitem{BPP1}  S.M. Bilenky, S. Pascoli and S.T. Petcov,
Phys. Rev. D {\bf 64} (2001) 053010;
S.T. Petcov, Phys. Scripta T {\bf 121} (2005) 94;
S. Pascoli and S. T. Petcov,
Phys. Rev. D {\bf 77} (2008) 113003. 



\bibitem{bb0nudata} See, e.g. C. Aalseth {\it et al.}, 
arXiv:hep-ph/0412300;
C.E. Aalseth {\it et al.} [IGEX Collaboration],
Phys. Atomic Nuclei {\bf 63} (2000) 1225;
H.~V.~Klapdor-Kleingrothaus {\it et al.},  
Phys. Lett. B {\bf 586} (2004) 198;
A.~S.~Barabash  [NEMO Collaboration],
arXiv:0807.2336 [nucl-ex];
C.~Arnaboldi {\it et al.}  [CUORICINO Collaboration],
Phys.\ Rev.\  C {\bf 78} (2008) 035502.


 \bibitem{HanPasc09} A. Atre {\it et al.},
JHEP {\bf 0905} (2009) 030.


  
\bibitem{del Aguila:2006dx}
  F.~del Aguila, J.~A.~Aguilar-Saavedra and R.~Pittau,
  J.\ Phys.\ Conf.\ Ser.\  {\bf 53} (2006) 506
  [arXiv:hep-ph/0606198].
  
  
\bibitem{FernandezMartinez:2007ms}
  E.~Fernandez-Martinez {\it et al.},
  Phys.\ Lett.\  B {\bf 649} (2007) 427.

\bibitem{Abada:2007ux}
  A.~Abada {\it et al.},
  JHEP {\bf 0712} (2007) 061.


\bibitem{Antusch:2008tz}
  S.~Antusch, J.~P.~Baumann and E.~Fernandez-Martinez,
  Nucl.\ Phys.\  B {\bf 810} (2009) 369.

\bibitem{Antusch:2006vwa}
  S.~Antusch {\it et al.},
  JHEP {\bf 0610} (2006) 084.
  
\bibitem{Merle:2006du}
  A.~Merle and W.~Rodejohann,
  Phys.\ Rev.\  D {\bf 73} (2006) 073012.

\bibitem{HPR83} A. Halprin, S.T. Petcov and S.P. Rosen, 
Phys. Lett. B {\bf 125} (1983) 335.

\bibitem{JV83} J. Vergados, Nucl. Phys. B {\bf 218} (1983) 109.

\bibitem{HaxStev84} W.C. Haxton and J. Stephenson,
Prog. Part. Nucl. Phys. {\bf 12} (1984) 409.


\bibitem{Blen10} M. Blennow {\it et al.}, 
arXiv:1005.3240.

\bibitem{Hirsch01} H. Paes {\it et al.},
Phys. Lett. B {\bf 498} (2001) 35.



\bibitem{Kersten:2007vk}
  J.~Kersten and A.~Y.~Smirnov,
  Phys.\ Rev.\  D {\bf 76} (2007) 073005.

\bibitem{Xing:2009ce}
  Z.~z.~Xing,
  Phys.\ Lett.\  B {\bf 679} (2009) 255.
  
\bibitem{Casas:2001sr}
  J.~A.~Casas and A.~Ibarra,
  Nucl.\ Phys.\  B {\bf 618} (2001) 171.

\bibitem{LeungSTP83} C.N. Leung and S.T. Petcov, 
Phys. Lett. B {\bf 125} (1983) 461.

\bibitem{LWDW83} 
D.~Wyler and L.~Wolfenstein,
Nucl. Phys.  B {\bf 218} (1983) 205.

\bibitem{Branco:1988ex}
  G.~C.~Branco, W.~Grimus and L.~Lavoura,
  Nucl.\ Phys.\  B {\bf 312} (1989) 492.
  
\bibitem{Amit10} R. Adhikari and A. Raychaudhuri, arXiv:1004.5111.


\bibitem{LW81} L. Wolfenstein, Nucl. Phys. B {\bf 186} (1981) 147.

\bibitem{STPPD82} S.T. Petcov, Phys. Lett. B {\bf 110} (1982) 245.


\bibitem{delAguila:2008hw}
  F.~del Aguila and J.~A.~Aguilar-Saavedra,
  Phys.\ Lett.\  B {\bf 672} (2009) 158
  [arXiv:0809.2096 [hep-ph]].

\bibitem{SKhalil10}  K. Huitu {\it et al.},
Phys. Rev. Lett. {\bf 101} (2008) 181802. 

\bibitem{SPTosh84} S.T. Petcov and 
S. Toshev, Phys. Lett. B {\bf 143} (1984) 175; 
K. Babu and E. Ma, Phys. Rev. Lett. {\bf 61} (1988) 674.

 
\bibitem{Dudas:2002ry}
  E.~Dudas and C.~A.~Savoy,
  Acta Phys.\ Polon.\  B {\bf 33} (2002) 2547.

\bibitem{Strum08} R. Franceschini, T. Hambye and A. Strumia,
arXiv:0805.1613.

 
\bibitem{Bonnet:2009ej}
  F.~Bonnet, D.~Hernandez, T.~Ota and W.~Winter,
  JHEP {\bf 0910} (2009) 076.
  

  



   
 \end{thebibliography}
\end{document}